\documentclass[a4paper,11pt]{article}
\pdfoutput=1 
\usepackage{jheppub} 
\usepackage{amsmath,amsfonts,amssymb,mathrsfs,graphicx,color,longtable,bm,wasysym}
\usepackage{hyperref}
\usepackage{graphicx}
\usepackage{array}
\usepackage{color}
\usepackage[usenames,dvipsnames,table]{xcolor}
\usepackage{tikz}
\usepackage[utf8]{inputenc}
\usepackage{epsfig,relsize,xspace}
\usepackage[T1]{fontenc}
\usepackage{bbold}
\usepackage{float}
\usepackage{amsmath}
\usepackage{empheq}
\usepackage{booktabs}
\usepackage{color}
\usepackage[utf8]{inputenc}
\usepackage{xspace}
\usepackage{scalerel}
\usepackage[most]{tcolorbox}
\usepackage{multirow}
\usepackage{scalefnt}
\usepackage{bold-extra}
\usepackage[shortlabels]{enumitem}
\usepackage[tikz]{bclogo}
\usepackage{subcaption}
\usepackage{tikz-feynman}
\usepackage{cancel}
\usepackage{verbatim}
\usetikzlibrary{arrows,shapes}
\usepackage{afterpage}
\usepackage{graphicx,grffile}

\allowdisplaybreaks

\def\TeV{\ifmmode {\mathrm{\ Te\kern -0.1em V}}\else
                   \textrm{Te\kern -0.1em V}\fi}
\def\GeV{\ifmmode {\mathrm{\ Ge\kern -0.1em V}}\else
                   \textrm{Ge\kern -0.1em V}\fi}
 
\definecolor{nicered}{rgb}{0.7,0.1,0.1}
\definecolor{nicegreen}{rgb}{0.1,0.5,0.1}
\definecolor{niceblue}{rgb}{0.0,0.1,0.7}
\hypersetup{colorlinks,citecolor=nicegreen,linkcolor=nicered,urlcolor=niceblue}
                                      
\def\bm#1{\mbox{\boldmath$#1$\unboldmath}}
\def \beq{\begin{equation}}
\def \eeq{\end{equation}}
\def \bea{\begin{eqnarray}}
\def \eea{\end{eqnarray}}

\begin{document}

\title{Beautiful and charming chromodipole moments}

\author[1]{Ulrich Haisch}

\author[1]{and Gabri{\"e}l Koole}

\affiliation[1]{Max Planck Institute for Physics, F{\"o}hringer Ring 6,  80805 M{\"u}nchen, Germany}

\emailAdd{haisch@mpp.mpg.de}
\emailAdd{koole@mpp.mpg.de}

\abstract{In the context of the Standard Model effective field theory we derive direct and indirect bounds on chromodipole operators involving the bottom and charm quark. We find that the experimental upper limit on the neutron electric dipole moment puts severe constraints on the imaginary parts of the Wilson coefficients of both  chromodipole operators. The magnitudes of the Wilson coefficients are instead only weakly constrained by dijet searches and $Z$-boson production in association with bottom-quark jets. Flavour physics does not provide meaningful bounds.}

\maketitle

\section{Introduction}
\label{sec:introduction}

Bottom quarks play an important role in the physics of the Standard Model~(SM) Higgs boson since ${\rm BR} \hspace{0.5mm} ( h \to b \bar b)$ is with around  58\% the  largest  Higgs  branching ratio. Due to the large backgrounds from multi-jet production in the dominant gluon-gluon fusion Higgs production mode, the most sensitive production channels for detecting $h \to b \bar b$ decays are the associated production of a Higgs boson and a $W$ or $Z$ boson ($Vh$), where the leptonic decay of the vector boson enables efficient triggering. In fact, the $h \to b \bar b$ decay mode has  been observed by both the ATLAS and the CMS collaborations at LHC Run~II~\cite{Aaboud:2018zhk,Sirunyan:2018kst}, and these measurements  constrain the $b \bar b$ signal strength in $Vh$ production to be SM-like within about~$25\%$ at the level of one standard deviation. 

Beyond the SM~(BSM) physics modifying the bottom-Higgs dynamics is therefore only weakly constrained by existing LHC measurements with even looser limits applying in the charm-Higgs case~\cite{Bishara:2016jga,Soreq:2016rae,Aaboud:2018fhh,Sirunyan:2018sgc,Sirunyan:2019qia,ATLAS-CONF-2019-029,ATLAS:2020wny}. In~the context of the SM effective field theory~(SMEFT) the dimension-six mixed-chirality operators  that can lead to modifications of the $h \to b \bar b$ and   $h \to c \bar c$ partial decay widths are either of Yukawa  or dipole type. While the Yukawa-type operators change the rates at leading order (LO) in QCD, the chromodipole operators 
\beq \label{eq:dipoleoperators}
Q_{bG} =    g_s \hspace{0.5mm} \bar q_3  \sigma_{\mu \nu}  T^a \hspace{0.25mm}  d_3 \hspace{0.25mm}  H \hspace{0.25mm}  G^{a,  \hspace{0.25mm} \mu \nu} \,, \qquad 
Q_{cG} =    g_s \hspace{0.5mm} \bar q _2 \sigma_{\mu \nu}  T^a \hspace{0.25mm}  u_2 \hspace{0.25mm}  \widetilde H \hspace{0.25mm}  G^{a,  \hspace{0.25mm} \mu \nu} \,, 
\eeq
start to contribute at the next-to-leading order~(NLO) level~\cite{Gauld:2016kuu}. Here $g_s$ denotes the QCD coupling,~$T^a$ are the $S U(3)$ colour generators and $G_{\mu \nu}^a$ is the gluon field strength tensor. The symbol~$q_f$ denotes left-handed quark doublets of flavour $f$, while $u_f$ and $d_f$ are the right-handed up- and down-type quark singlets, $\sigma_{\mu \nu} = i/2 \hspace{0.5mm} (\gamma_\mu \gamma_\nu - \gamma_\nu \gamma_\mu)$ with $\gamma_\mu$ the usual Dirac matrices, $H$ denotes the SM Higgs doublet and the shorthand notation $\widetilde H_i = \epsilon_{ij} \hspace{0.5mm} \big ( H_j \big ) ^\ast$   with $\epsilon_{ij}$ totally antisymmetric and~$\epsilon_{12} =1$ has been used.

In contrast to the SMEFT Yukawa-type operators that can only be constrained by processes involving a Higgs or Higgses both real or virtual, the operators~(\ref{eq:dipoleoperators}) can also be bounded by measurements of observables that involve bottom (charm) quarks and possibly other particles but no Higgs boson. The goal of this work is to derive bounds on the Wilson coefficients of the operators~$Q_{bG}$ and $Q_{cG}$ using existing measurements of both low- and high-energy observables of the latter type ---  the~high-luminosity LHC~(HL-LHC) constraints on~(\ref{eq:dipoleoperators}) from Higgs physics can be found in the publications~\cite{Hayreter:2013kba,Bramante:2014hua} and will be studied elsewhere. 

Our article is organised as follows. In Section~\ref{sec:dijets} we derive the present constraints from LHC~Run~II searches of unflavoured jet pairs, while Section~\ref{sec:bjets} contains a discussion of the bounds that the latest  LHC searches for bottom-quark jets ($b$-jets) provide (cf.~also~\cite{Bramante:2014hua} for an earlier study). The constraints that we derive from $Z$-boson production in association with~$b$-jets are discussed in Section~\ref{sec:Zbjets}. In Section~\ref{sec:flavour} and  Section~\ref{sec:nEDM} we present the limits from flavour physics and the neutron electric dipole moment~(nEDM), respectively. We~summarise our main results and give a brief discussion of the implications of our findings for explicit BSM models  in Section~\ref{sec:discussion}.

\section{Constraints from dijet angular distributions}
\label{sec:dijets}

\begin{figure}[!t]
\begin{center}
\begin{subfigure}[b]{0.5\linewidth}
      \centering
       \scalebox{1.2}{
      \begin{tikzpicture}
      \begin{feynman}
      \vertex (a1) {\(g\)};
       \vertex[below=1.5cm of a1] (a2){\(g\)};
       \vertex[below=0.75cm of a1] (a3);
       \vertex[right=1.3cm of a3] (a4);
       \vertex[right=1.cm of a4] (a5);
       \vertex[right=3.6cm of a1](a6) {\(b\)};
       \vertex[right=3.6cm of a2] (a7){\(\bar b\)};     
       \diagram* {
         {[edges=gluon]
           (a1)--(a4)--(a2),},
           (a4) -- [gluon, edge label=\(g\)] (a5),
         {[edges=anti fermion]
           (a6)--(a5)--(a7),},
         };
         \vertex[dot,fill=black] (d) at (a4){};
       \vertex[square dot,fill=nicered] (d) at (a5){};
      \end{feynman}
      \end{tikzpicture}
      }
\end{subfigure}%
\begin{subfigure}[b]{0.5\linewidth}
       \centering
        \scalebox{1.2}{
       \begin{tikzpicture}
       \begin{feynman}
          \vertex (a1) {\(g\)};
          \vertex[below=1.5cm of a1] (a2){\( g\)};
          \vertex[right=1.3cm of a1] (a3);
          \vertex[right=1.3cm of a2] (a4);
          \vertex[right=1.1cm of a3] (a5){\( b\)};
          \vertex[right=1.1cm of a4] (a6){\(\bar b\)}; 
          \vertex[below=0.75 of a3](a7);       
          \diagram* {
            (a1)--[gluon](a7)--[gluon](a2),
            (a7)--[fermion] (a5),
            (a7)--[anti fermion] (a6),
          };
          \vertex[square dot,fill=nicered] (d) at (a7){};
        \end{feynman}
       \end{tikzpicture}
       }
\end{subfigure}
\end{center}
\begin{center}
\begin{subfigure}[b]{0.5\linewidth}
      \centering
       \scalebox{1.2}{
      \begin{tikzpicture}
      \begin{feynman}
          \vertex (a1) {\(g\)};
          \vertex[below=1.6cm of a1] (a2){\( b\)};
          \vertex[right=1.7cm of a1] (a3);
          \vertex[right=1.7cm of a2] (a4);
          \vertex[right=1.3cm of a3] (a5){\( b\)};
          \vertex[right=1.3cm of a4] (a6){\(g\)}; 
          \vertex[below=0.3cm of a3] (a7);
          \vertex[above=0.3cm of a4] (a8);            
          \diagram* {
            (a1)--[gluon](a7)--[anti fermion](a8)--[anti fermion](a2),
            (a7)--[fermion] (a5),
            (a8)--[gluon] (a6),
          };
           \vertex[dot,fill=black] (d) at (a8){};
           \vertex[square dot,fill=nicered] (d) at (a7){};
        \end{feynman}      
        \end{tikzpicture}
        }
\end{subfigure}%
\begin{subfigure}[b]{0.5\linewidth}
        \centering
         \scalebox{1.2}{
        \begin{tikzpicture}
        \begin{feynman}
          \vertex (a1) {\(g\)};
          \vertex[below=1.5cm of a1] (a2){\( b\)};
          \vertex[right=1.3cm of a1] (a3);
          \vertex[right=1.3cm of a2] (a4);
          \vertex[right=1.1cm of a3] (a5){\( b\)};
          \vertex[right=1.1cm of a4] (a6){\(g\)}; 
          \vertex[below=0.75 of a3](a7);       
          \diagram* {
            (a1)--[gluon](a7)--[anti fermion](a2),
            (a7)--[fermion] (a5),
            (a7)--[gluon] (a6),
          };
          \vertex[square dot,fill=nicered] (d) at (a7){};
        \end{feynman}
       \end{tikzpicture}
        }
\end{subfigure}
\vspace{-2mm}
\caption{\label{fig:dijets} Tree-level diagrams leading to a correction to dijet production. The red squares denote the insertion of the bottom-quark chromomagnetic dipole moment.} 
\end{center}
\end{figure}
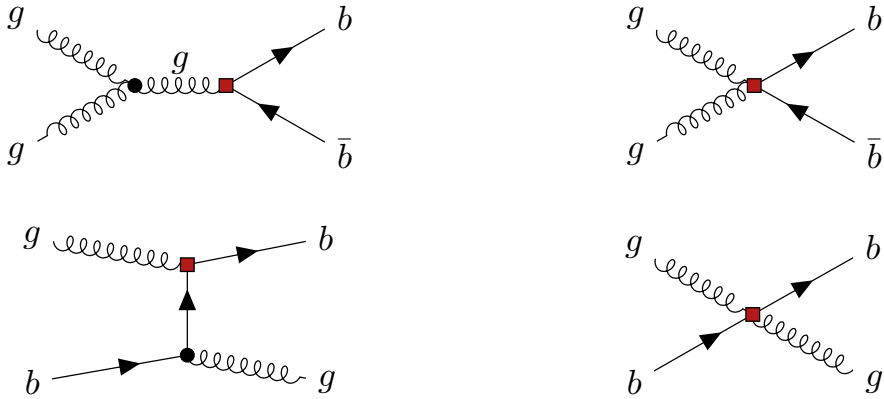

Jet physics has previously been  used to put constraints on SMEFT operators~\cite{Krauss:2016ely,Alioli:2017jdo,Alte:2017pme,Hirschi:2018etq,Keilmann:2019cbp,Goldouzian:2020wdq}. In~this section we will exploit  searches for unflavoured jet pairs to constrain the magnitudes of the Wilson coefficients of the operators introduced in~(\ref{eq:dipoleoperators}).  For the case of the  bottom-quark chromodipole operator~$Q_{bG}$, examples of contributing Feynman diagrams are shown in~Figure~\ref{fig:dijets}. The relevant collider searches look either for resonances in the dijet spectrum or analyse the angular distribution of dijet production (see~\cite{Sirunyan:2018wcm,Aad:2019hjw,Sirunyan:2019vgj,Aad:2020cws} for recent LHC results). 

The quantity of interest in our case  is the jet angular distribution,~i.e.~the differential cross section for a pair of jets with invariant mass~$M_{jj}$ produced at an angle $\hat \theta$ to~the beam direction in the jet-jet centre-of-mass~(CM) frame. Compared to resonance searches, the dijet angular distribution has the salient advantage that it also allows to constrain broad $s$-channel resonances or modifications in the spectrum due to the presence of higher-dimensional operators in a rather model-independent fashion. This is due to the fact that the dominant channels in QCD dijet production have the familiar Rutherford scattering behaviour $d\sigma/d\cos \hat \theta \propto 1/\sin^4 \hspace{0.25mm} \big ( \hat \theta/2 \big )$ at small angle $\hat \theta$, which is characteristic for $t$-channel exchange of a massless spin-one boson. In order to remove the Rutherford singularity, one usually considers the dijet cross sections differential in
\beq \label{eq:chi}
\chi = \frac{1 +\cos \hat \theta}{1 - \cos \hat \theta} \,.
\eeq
In the small angle limit,~i.e.~$\chi \to \infty$, the partonic differential QCD cross section then behaves as $d\sigma/d\chi \propto \text{const}$. Relative to the QCD background, the production of a heavy resonance or an effective operator leads to additional hard scattering and hence more jets perpendicular to the beam. In turn one expects a deviation from the QCD prediction in form of an enhanced activity of high-energetic jets in the central region of the detector. If the angular distributions receive contributions from the presence of a heavy degree of freedom or an operator, one should see an excess of events in $d\sigma/d\chi$  for $\chi \to 1$ and large~$M_{jj}$ with respect to the (almost) flat QCD spectrum.

\begin{figure}[!t]
\begin{center}
\includegraphics[width=0.475\textwidth]{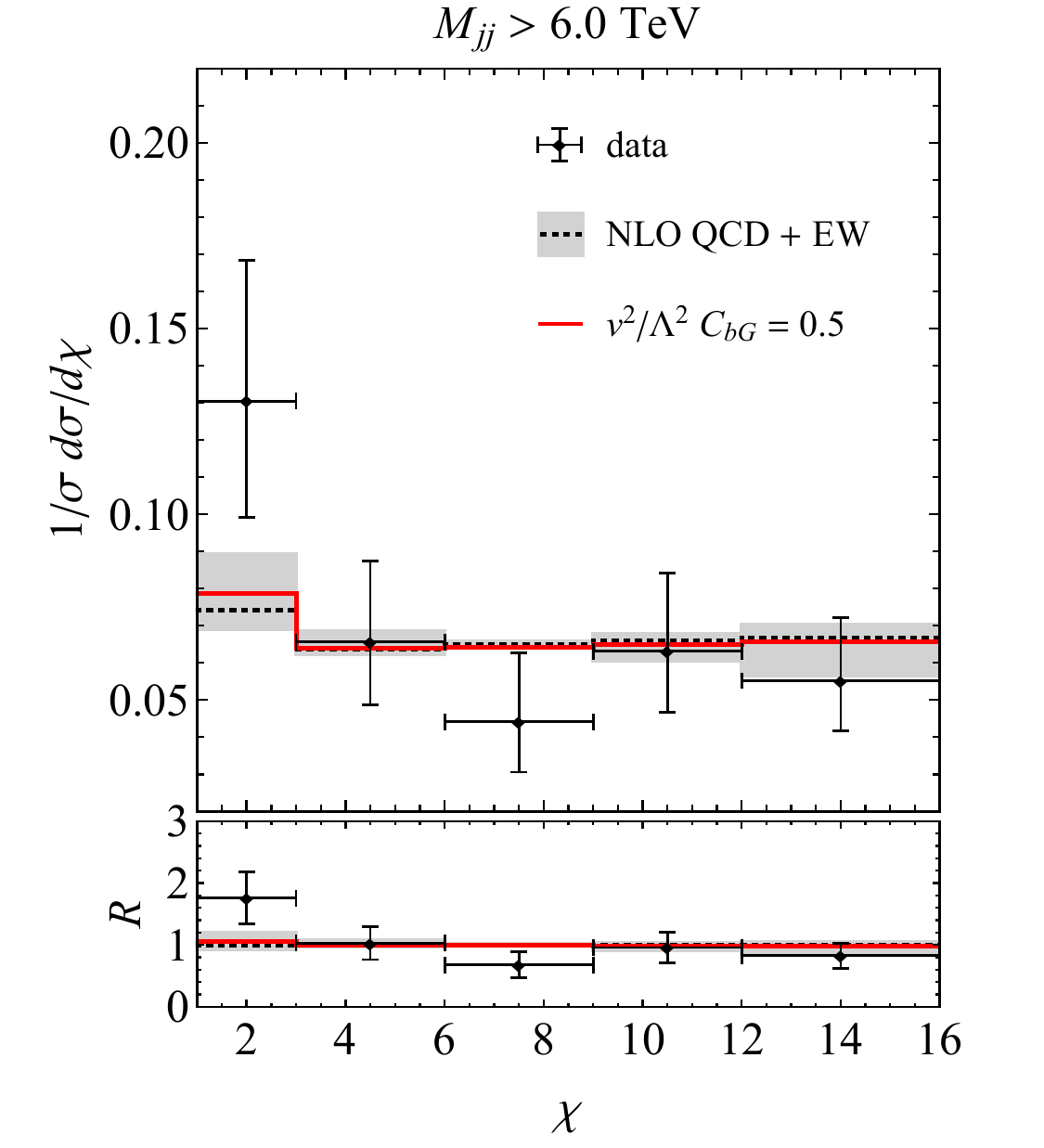} \quad 
\includegraphics[width=0.475\textwidth]{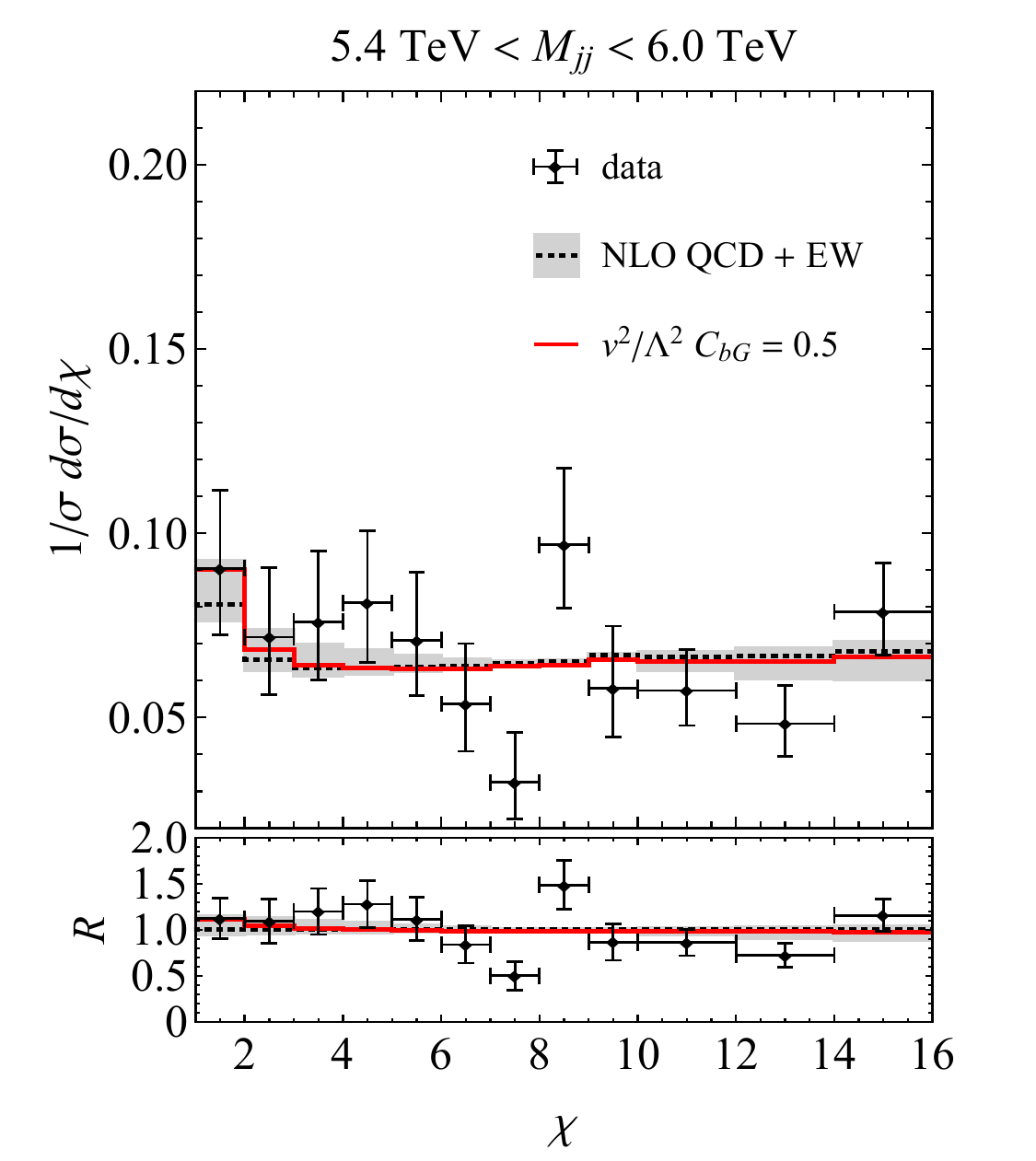}

\vspace{2mm}

\includegraphics[width=0.475\textwidth]{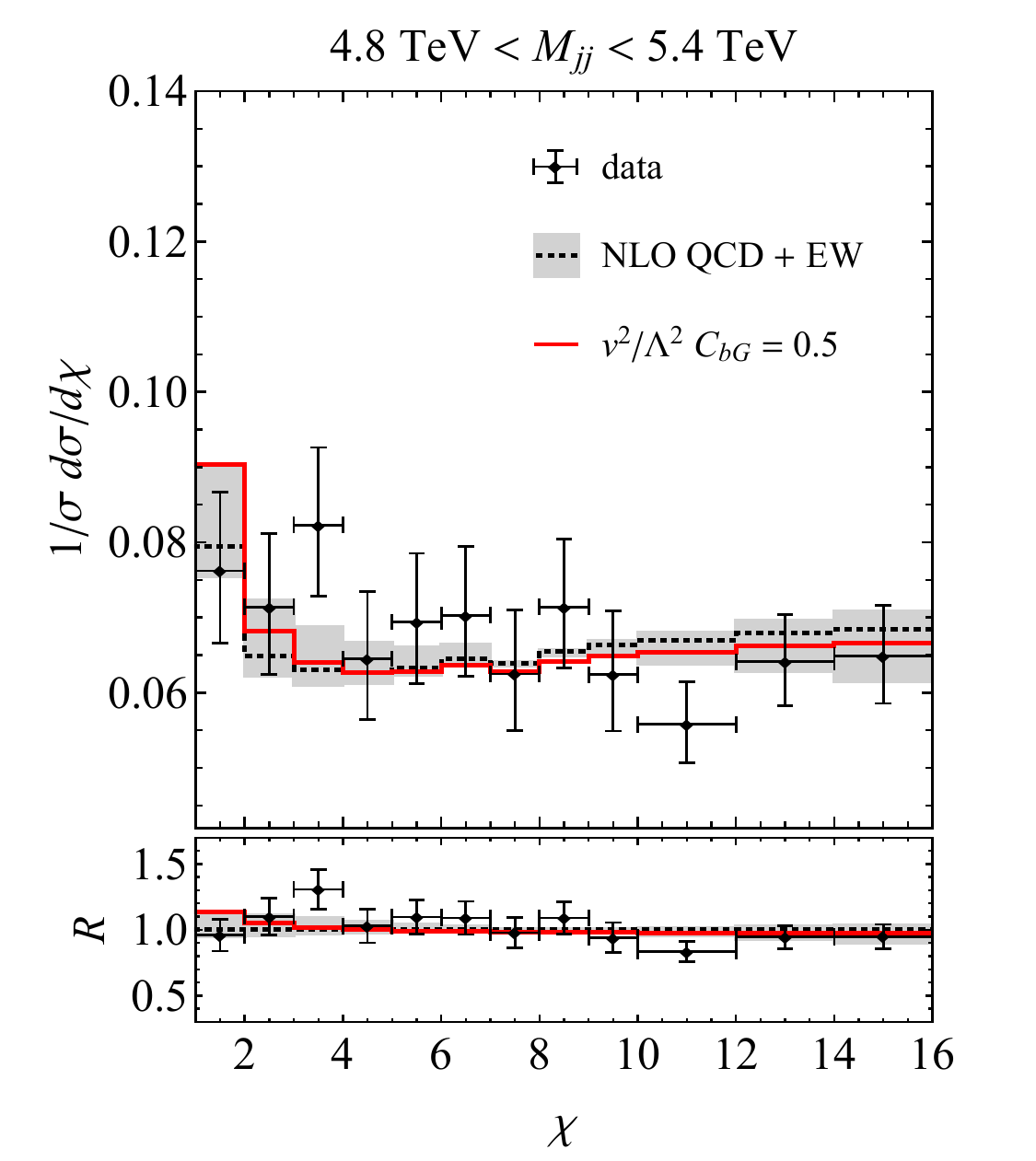}
\vspace{0mm} 
\caption{\label{fig:dijetspectrum1} Normalised $\chi$ distribution in the three highest mass bins studied in the CMS analysis~\cite{Sirunyan:2018wcm}. Unfolded data from~\cite{Sirunyan:2018wcm} are compared to the SM prediction including NLO QCD and EW corrections (black dotted line). The error bars represent statistical and experimental systematic uncertainties combined in quadrature. Theoretical uncertainties are indicated as a grey band. Also shown is the prediction for $v^2/\Lambda^2 \hspace{0.5mm} C_{bG} = 0.5$ (red line). The lower panels show the ratio of the unfolded data distributions and the NLO predictions as well as the new-physics distributions. See~main text for more details.} 
\end{center}
\end{figure}

\begin{figure}[!t]
\begin{center}
\includegraphics[width=0.475\textwidth]{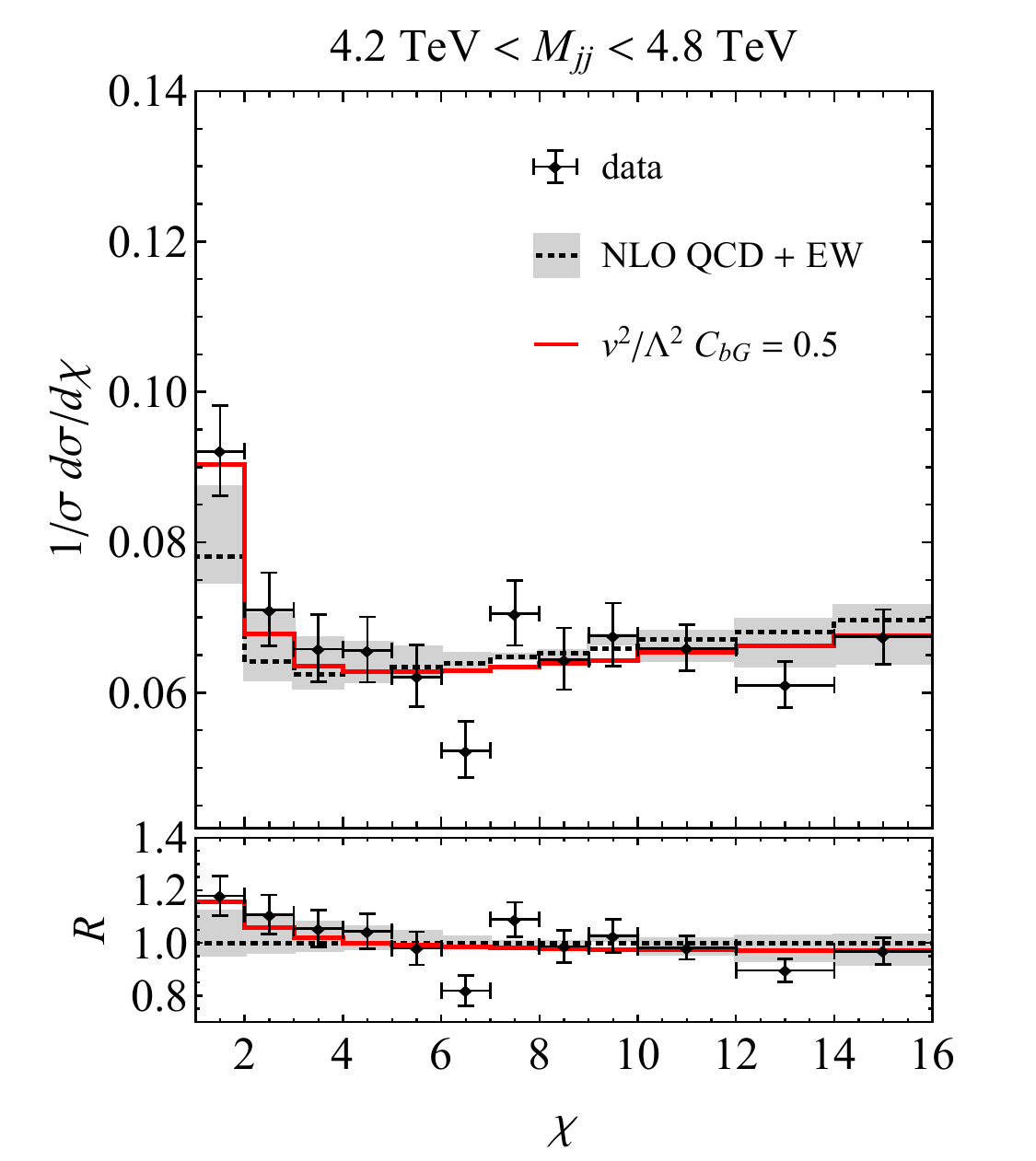} \quad 
\includegraphics[width=0.475\textwidth]{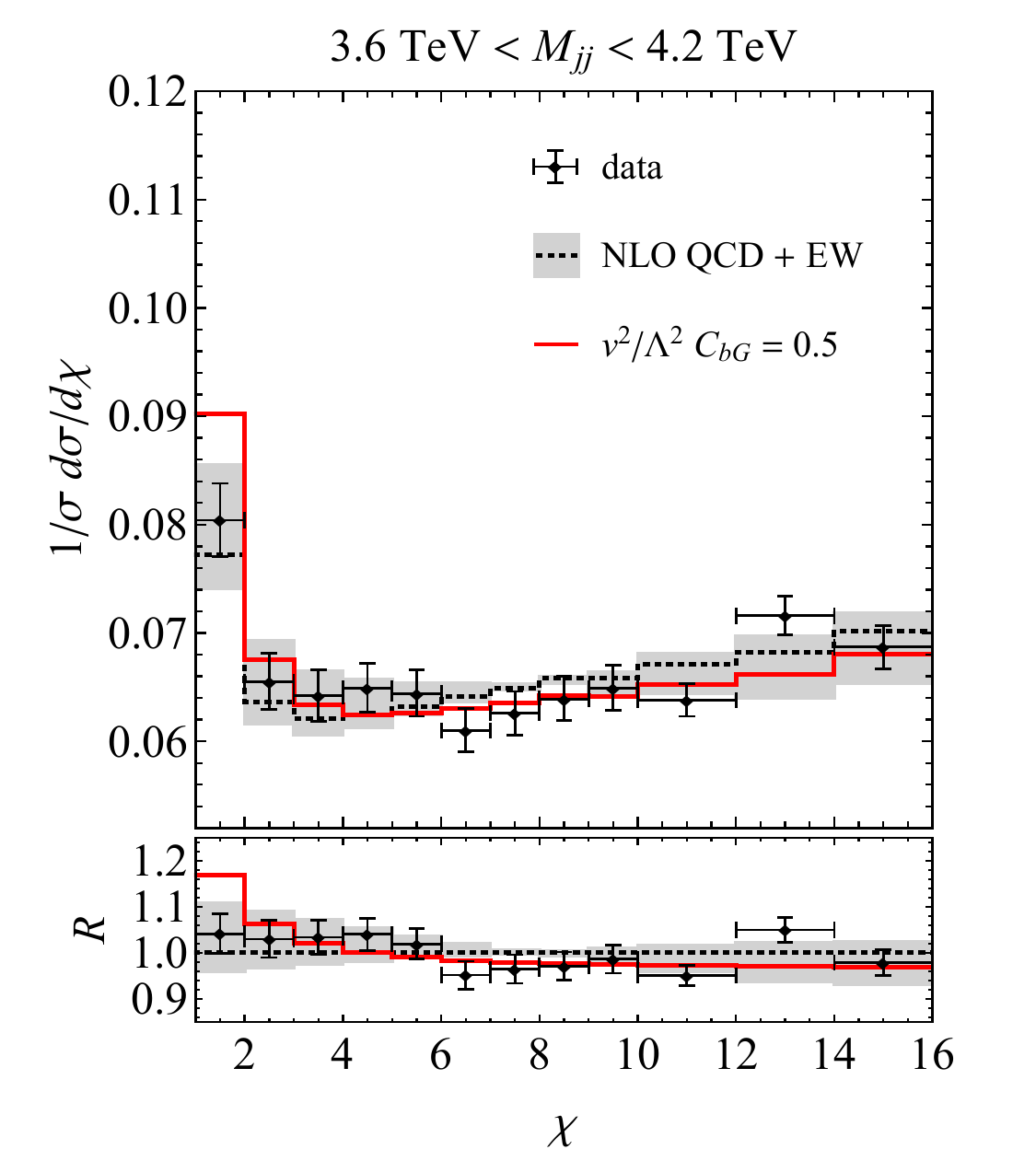}

\vspace{2mm}

\includegraphics[width=0.475\textwidth]{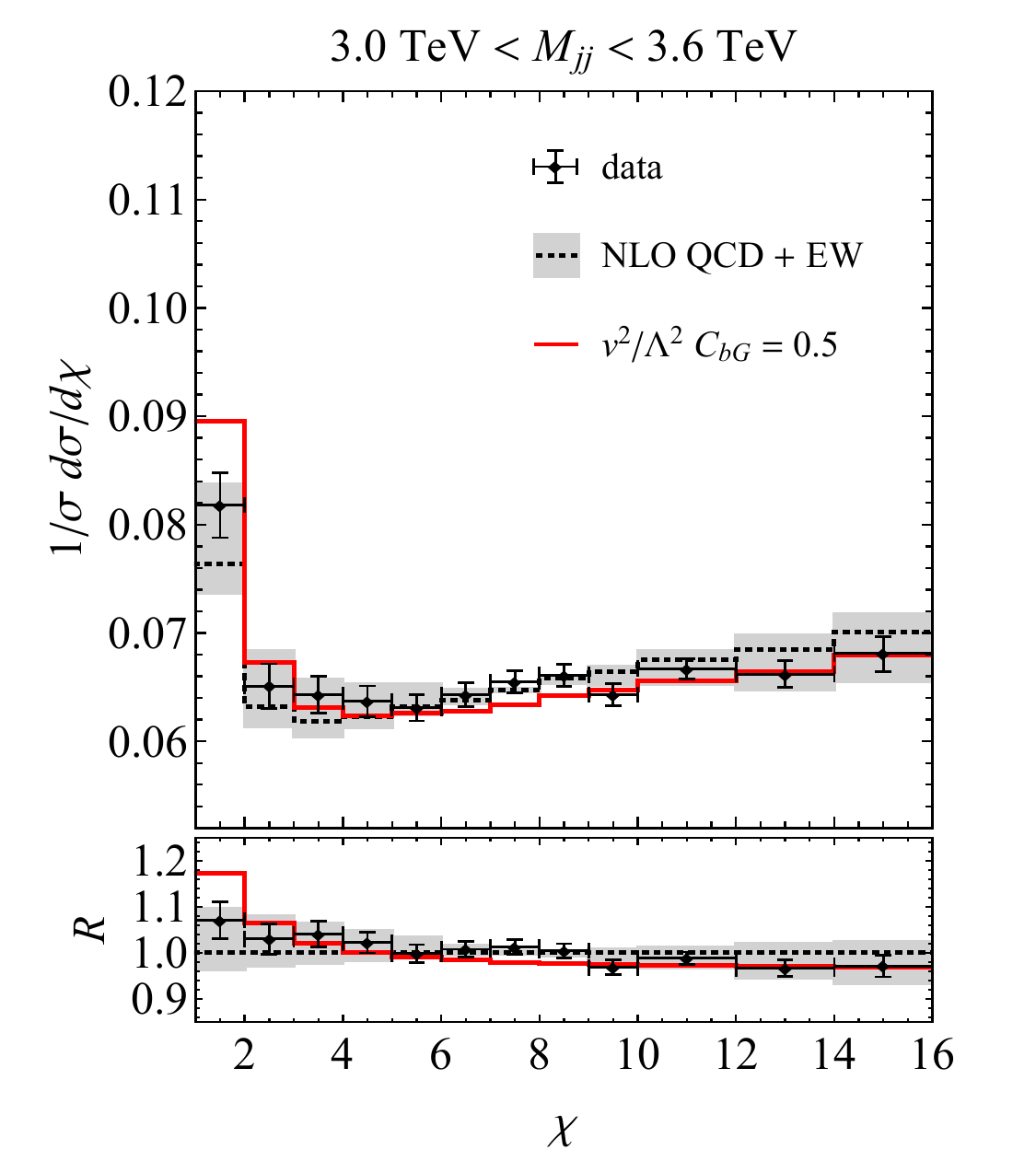} \quad 
\includegraphics[width=0.475\textwidth]{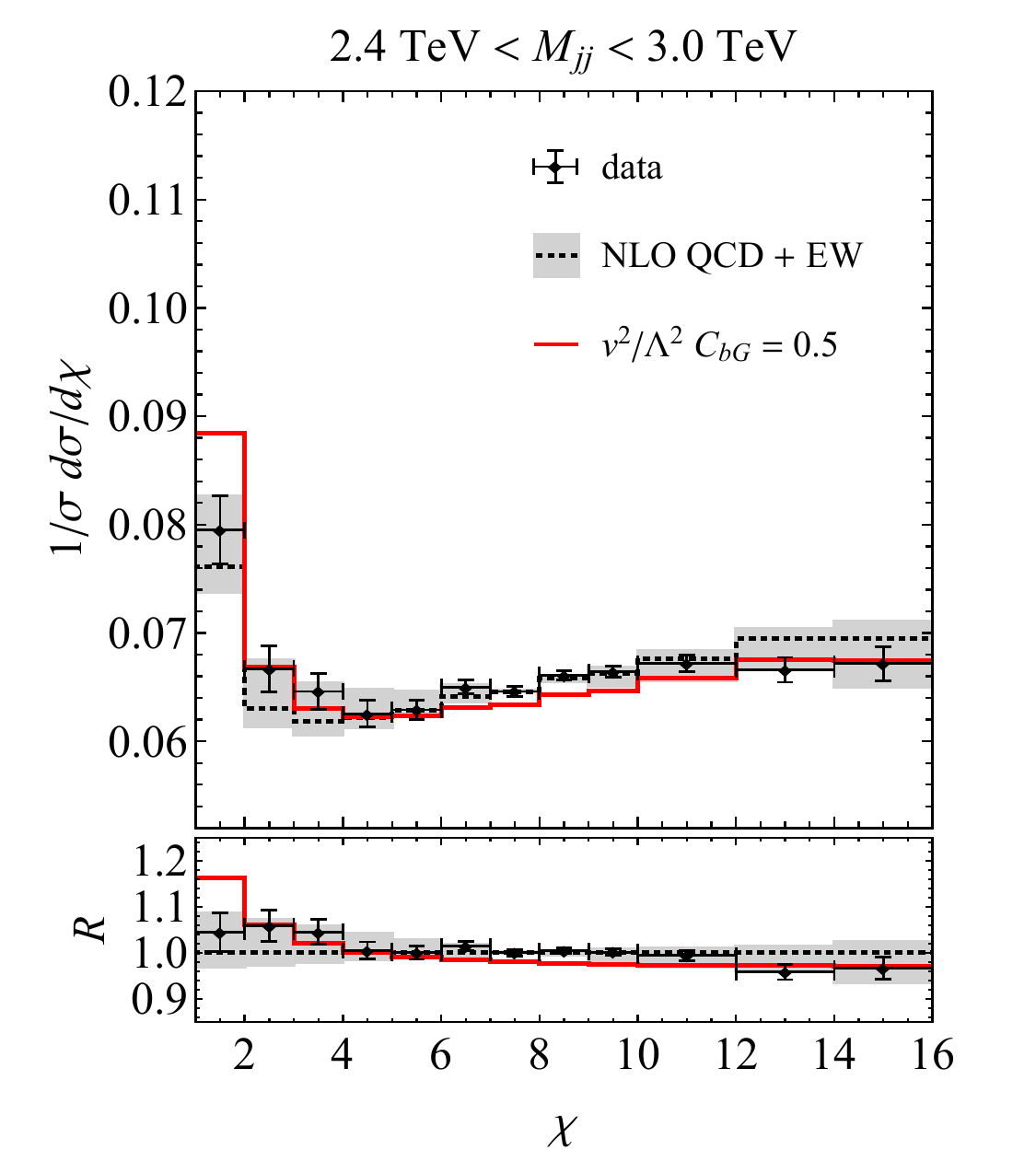}
\vspace{0mm} 
\caption{\label{fig:dijetspectrum2} As Figure~\ref{fig:dijetspectrum1} but for the four lower mass bins considered in the CMS analysis~\cite{Sirunyan:2018wcm}.} 
\end{center}
\end{figure}

By calculating the partonic differential cross section for the interference of the effective operator $Q_{bG}$ with the SM and with itself, we find the following expressions
\bea \label{eq:jetjetQbG}
\begin{split}
\left ( \frac{d\sigma (gg \to b \bar b)}{d \chi} \right )_{bG} & = \frac{\pi \hspace{0.25mm} \alpha_s^2}{2 \hspace{0.25mm}  M_{jj}^2} \left [ \frac{ 4-\chi+4\hspace{0.25mm}\chi^2 }{3 \hspace{0.25mm} \chi \left ( 1 +\chi \right)^2}  \hspace{0.25mm}  \frac{v^2}{\Lambda^2} \hspace{0.5mm}  y_b \hspace{0.5mm}  {\rm Re} \hspace{0.25mm} \big ( C_{bG} \big )    + \frac{7 \hspace{0.25mm} M_{jj}^2}{3 \left ( 1 +\chi \right)^2} \hspace{0.25mm} \frac{v^2}{\Lambda^4} \hspace{0.25mm} \left | C_{bG}   \right |^2 \right ] \,, \\[2mm]
\left ( \frac{d\sigma (gb \to g b)}{d \chi} \right )_{bG} & =  \frac{\pi \hspace{0.25mm} \alpha_s^2}{2 \hspace{0.25mm}  M_{jj}^2} \left [  \frac{8 \left (4 + 9 \hspace{0.25mm} \chi+9\hspace{0.25mm}\chi^2 \right )}{9 \hspace{0.25mm} \chi   \left ( 1 +\chi \right)^3}  \hspace{0.25mm}  \frac{v^2}{\Lambda^2} \hspace{0.5mm}  y_b \hspace{0.5mm}  {\rm Re} \hspace{0.25mm} \big ( C_{bG} \big )    + \frac{56\hspace{0.5mm}  M_{jj}^2}{9 \left ( 1 +\chi \right)^3} \hspace{0.25mm} \frac{v^2}{\Lambda^4} \hspace{0.25mm} \left | C_{bG}   \right |^2 \right ]  \,, \hspace{4mm}
\end{split}
\eea
where $\alpha_s = g_s^2/(4 \pi)$ is the strong coupling constant, $y_b = \sqrt{2} m_b/v$ is the bottom-quark Yukawa coupling with $m_b$ the bottom-quark mass and $v \simeq 246 \, {\rm GeV}$  the vacuum expectation value of the Higgs field. The result for the differential cross section for the process $g \bar b \to g \bar b$ is identical to the one for $g  b \to g b$, and the results for the case of the operator $Q_{cG}$  are simply obtained from the above expressions by replacing~$b$ everywhere by~$c$. Notice that the first terms in~(\ref{eq:jetjetQbG}) which arise from the interference with the SM are suppressed by one power of $y_b$ which provides the chirality flip needed to obtain a non-zero result. The second terms which are due to the interference of the operator $Q_{bG}$ with itself are instead enhanced by two powers of the jet-jet invariant mass $M_{jj}$. As a result the quadratic term will in practice always provide the dominant contribution to the dijet spectrum. Given that in addition the gluon-quark initiated channels are strongly suppressed by heavy-quark parton distribution functions~(PDFs) this means that the limit that we derive below holds to very good approximation for both $Q_{bG}$ and $Q_{cG}$. 

In terms of the partonic differential cross sections~(\ref{eq:jetjetQbG}), the contribution to the double differential cross section for a pair of jets with invariant mass $M_{jj}$  produced at $\chi$ that is due to $Q_{bG}$, can be written in the following way
\beq \label{eq:d2sigma}
\left ( \frac{d^2 \sigma}{d M_{jj}^2 \hspace{0.5mm} d \chi} \right )_{bG} =  \frac{M_{jj}^2}{s} \sum_{\{ i j \}} \, f\!\!f_{ij} \left (M_{jj}^2/s, \mu_F \right )  \, \left ( \frac{d \sigma (ij \to f )}{d \chi} \right )_{bG} \,,
\eeq
where 
\beq \label{eq:lumis}
 f\!\!f_{ij} \left (\tau, \mu_F \right ) = \frac{1}{s} \frac{2}{1+\delta_{ij}} \int_\tau^1  \! \frac{dx}{x} \, f_{i/p} (x,\mu_F)  \,  f_{j/p} (\tau/x,\mu_F) \,,
\eeq
are the so-called parton luminosities, the sum  runs over all pairs of incoming partons $\{ i j \}$ and $s$~denotes the squared CM energy of the collider.  The  parton luminosities are obtained from a convolution of the universal non-perturbative PDFs $f_{i/p} (x, \mu_F)$, which describe the probability of finding the parton~$i$ in the proton with longitudinal momentum fraction $x$. The variable $\mu_F$ that enters~(\ref{eq:d2sigma}) and~(\ref{eq:lumis}) denotes the factorisation scale. 

In Figures~\ref{fig:dijetspectrum1} and~\ref{fig:dijetspectrum2} the measured normalised $\chi$ distribution for the seven different mass bins unfolded to particle level are compared to the corresponding SM predictions including both QCD and  electroweak~(EW) corrections at the NLO. The data and the SM predictions are both taken from the CMS analysis~\cite{Sirunyan:2018wcm}. For  comparison the normalised~$\chi$ distributions assuming $v^2/\Lambda^2 \hspace{0.5mm} C_{bG} = 0.5$  are also shown. The new-physics distributions are obtained at~LO using {\tt CT14nnlo\_as\_0118} PDFs~\cite{Dulat:2015mca}  with renormalisation and factorisation scale set to~$M_{jj}$.  These PDFs have also been used in~\cite{Sirunyan:2018wcm} and are accessed through  {\tt ManeParse}~\cite{Clark:2016jgm}. The~normalised~$\chi$ distributions are then multiplied by a bin-wise $K$-factor  to obtain a reshaped spectrum that includes effects originating from NLO QCD and EW corrections,  additional non-perturbative QCD effects and the detector resolution. The $K$-factor is obtained by  calculating the ratio between the  central value of the CMS~SM prediction and our LO~SM prediction. Notice that applying the same rescaling factor to both the QCD and new-physics results is based on the assumption that the effects of the Monte~Carlo~(MC) shower and  the event detection depend only on the invariant mass $M_{jj}$ of the dijet final state, but not on the precise form of the new-physics signal. To determine to which extend this assumption is justified would require to perform a dedicated simulation of  including  NLO QCD and EW corrections,  parton showering and detector effects. However, such an analysis is beyond the scope of this work. 

To find the 95\%~confidence level~(CL)~limit on the magnitude of the Wilson coefficient of the dipole operator $Q_{bG}$, we perform a  $\chi^2$ fit to the normalised angular distribution in all seven dijet invariant mass bins,  including both experimental and theoretical uncertainties. Minimising the   $\chi^2$ and requiring $\Delta \chi^2 = \chi^2 - \chi^2_{\rm min} = 3.84$ leads to  
\beq \label{eq:dijetbound1}
\frac{\big | \hspace{0.125mm} C_{bG} \hspace{0.125mm} \big |}{\Lambda^2} < \left ( \frac{1}{380 \, {\rm GeV}} \right )^2 \,.
\eeq
The same bound applies approximately also  in the case of $Q_{cG}$. Notice that since the bounds~(\ref{eq:dipoleboundb}) and~(\ref{eq:dipoleboundc}) are significantly stronger than the latter limit, the exclusion~(\ref{eq:dijetbound1}) is in fact a bound on the real part of the Wilson coefficients~$C_{bG}$ and~$C_{cG}$.  We add that non-zero values $v^2/\Lambda^2 \hspace{0.5mm} |C_{bG}|\simeq 0.3$ of the Wilson coefficient are, in fact, preferred by the data, and that the most sensitive bins in our $\chi^2$ fit are  the three bins of~\cite{Sirunyan:2018wcm} that cover the range $2.4 \, {\rm TeV} < M_{jj} < 4.2 \, {\rm TeV}$ of invariant dijet masses. Notice finally that analyses of the jet-jet angular distribution in dijet production can also be used to put constraints on the chromodipole operators involving light quarks. Since the focus of this work lies on deriving constraints on the chromodipole operators involving bottom and charm quarks, we however do not entertain this possibility here.

\section{Constraints from searches for $\bm{b}$-jet pairs}
\label{sec:bjets}

In addition to using data for unflavoured dijet production, one can also exploit the measurements of mass distributions of jet pairs with one or two jets identified as $b$-jets~\cite{CMS-PAS-EXO-12-023,Aaboud:2018tqo,Sirunyan:2018ikr,Aad:2019hjw} to constrain the Wilson coefficient of the operator $Q_{bG}$ --- see also~\cite{Bramante:2014hua} for an earlier study. 

\begin{figure}[!t]
\begin{center}
\includegraphics[width=0.475\textwidth]{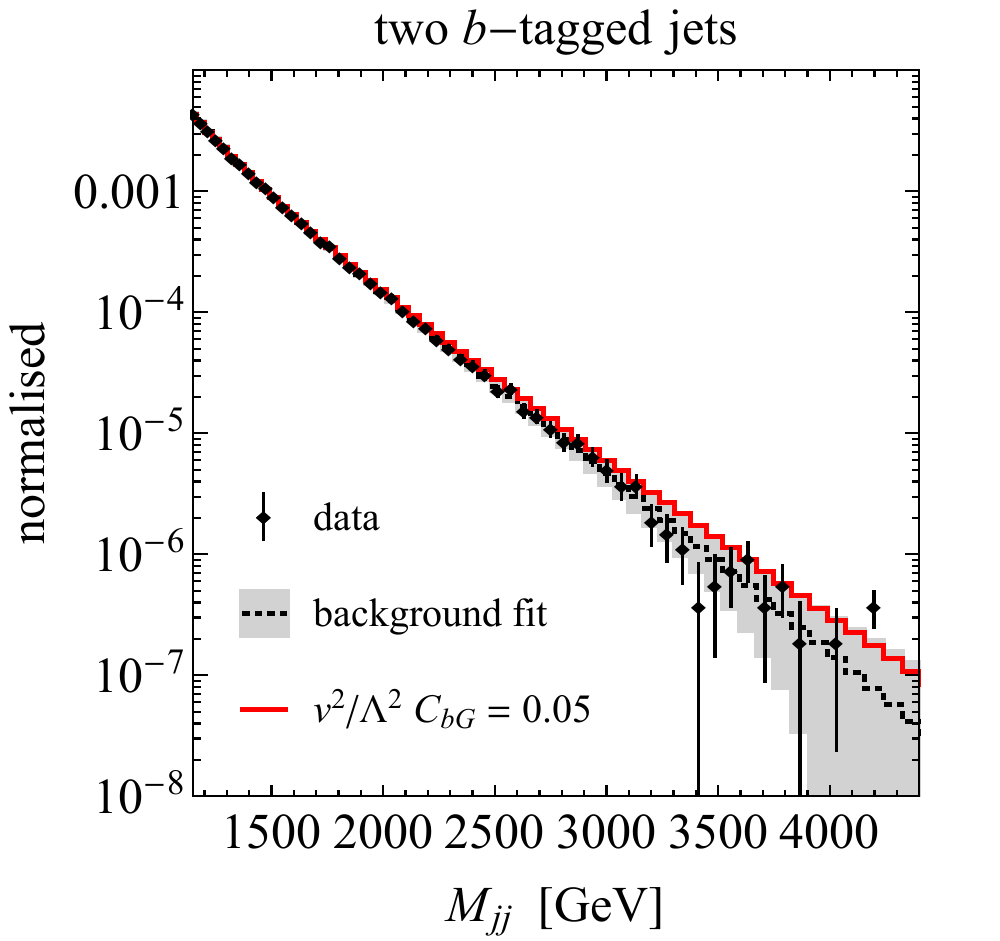} \quad 
\includegraphics[width=0.455\textwidth]{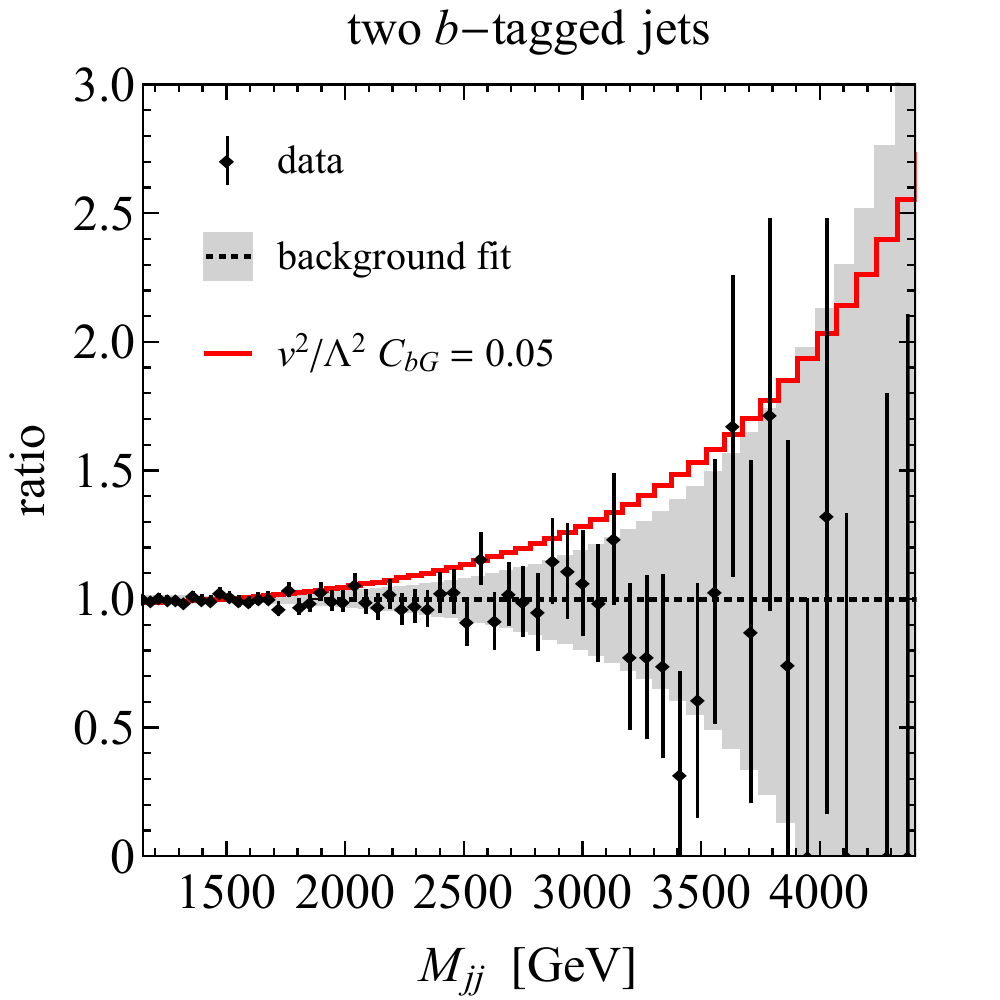}
\vspace{2mm} 
\caption{\label{fig:bjets} Left: Normalised dijet invariant mass distributions for the category with two $b$-jets as measured by ATLAS in~\cite{Aad:2019hjw}. The black dotted line is the central value of the background fit performed by ATLAS, the  grey band indicates the associated uncertainties and the red curve shows the new-physics prediction assuming $v^2/\Lambda^2 \hspace{0.5mm} C_{bG} = 0.05$. Right:~Ratio of the  data and the new-physics prediction for $v^2/\Lambda^2 \hspace{0.5mm} C_{bG} = 0.05$ to the central value of the background fit. The colour coding resembles the one used in the left panel. Consult the main text for further details.} 
\end{center}
\end{figure}

To calculate the differential cross sections for  the production of $b$-jet pairs we have used a {\tt FeynRules~2}~\cite{Alloul:2013bka}  implementation of the operator $Q_{bG}$ in the {\tt UFO} format~\cite{Degrande:2011ua}. The~generation and showering of the  samples has been   performed at LO  with {\tt MadGraph5\_aMC\@NLO}~\cite{Alwall:2014hca} and {\tt PYTHIA~8.2}~\cite{Sjostrand:2014zea}, respectively, using  {\tt NNPDF31\_nlo\_as\_0118} PDFs~\cite{Ball:2017nwa} and working in the four-flavour scheme.  The background distribution was corrected to the NLO prediction  using the matrix elements calculated in~\cite{Nason:1987xz} as implemented in {\tt MCFM}~\cite{Campbell:2019dru}. Hadronic jets~are built using the anti-$k_t$ algorithm~\cite{Cacciari:2008gp} with a radius parameter of $R=0.4$, as implemented in~{\tt FastJet}~\cite{Cacciari:2011ma}.  We furthermore rely on {\tt DELPHES~3}~\cite{deFavereau:2013fsa} as a fast detector simulation and on  {\tt CheckMATE~2}~\cite{Dercks:2016npn}. Our~event selection follows the ATLAS analysis~\cite{Aad:2019hjw} which is  based on $139 \, {\rm fb}^{-1}$ of dijet data collected at 13~TeV CM energy. We require at least two jets~($j$) with a transverse momentum $p_T(j)$  satisfying $p_T(j) > 150 \, {\rm GeV}$ and the azimuthal angle difference $\Delta \phi (j_1 j_2)$ between the two leading jets $j_1$ and  $j_2$ must fulfill $\left |\Delta \phi (j_1 j_2)  \right | > 1.0$. The  two leading jets must be $b$-tagged and their pseudorapidities must satisfy $\left |\eta (j) \right | < 2.0$. The  $b$-tagging algorithm is taken from the ATLAS publication~\cite{Aad:2019aic}, and in accordance with~\cite{Aad:2019hjw}   a $b$-tagging working point is chosen that yields a  $b$-tagging efficiency of~77\%,  a  $c$-jet rejection  of~5~and a light-flavour jet rejection  of~110. Furthermore, to suppress the QCD background a selection cut of $\left | y_\ast  \right | < 0.8$ is imposed, where  $y_\ast = \big (y (j_1) - y (j_2) \big )/2$ with $y (j_1) $ and $y(j_2) $ the rapidities of the leading and subleading jet, respectively.  Both the QCD background and the new-physics samples are generated binned in $p_T ( j)$ and the resulting dijet mass distributions are fit to the parametric function
\beq \label{eq:fdijet}
f(x)=p_1 \left (1 - x \right )^{\hspace{0.125mm} p_2} x^{\hspace{0.25mm} p_3+p_4 \ln x} \,,
\eeq
where $x = M_{jj}/\sqrt{s}$ and $p_i$ with $i = 1,2,3,4$  four fitting parameters. Given the data-driven background estimate, the ATLAS  measurement~\cite{Aad:2019hjw} is not a measurement of the  absolute cross section of dijet production, making it  insensitive to small  overall shifts in the $b \bar b$ production rates. The ATLAS analysis is however sensitive to the shape of the dijet mass distribution, which can be determined in normalised rates. 

In Figure~\ref{fig:bjets} we show the results for the normalised invariant dijet mass spectra requiring two $b$-jets. The black dotted line in the left plot denotes the central values of the background fit performed by ATLAS in~\cite{Aad:2019hjw}, while the red curve shows our new-physics prediction for $v^2/\Lambda^2 \hspace{0.5mm} C_{bG} = 0.05$. The grey band indicates the uncertainties associated to the background prediction as provided in~\cite{Aad:2019hjw}. On the right-hand side the ratio of the  data and the new-physics prediction to the central value of the background fit is displayed. One observes that the contributions associated to $Q_{bG}$ lead to an enhancement in the tail of the $M_{jj}$ distribution with respect  to the background  prediction. 

By performing a $\Delta \chi^2$ fit to the normalised spectra over the whole range of invariant dijet masses covered in the ATLAS study~\cite{Aad:2019hjw},~i.e.~$1133 \, {\rm GeV} < M_{jj} <  4595 \, {\rm GeV}$, we obtain the following 95\%~CL limit 
\beq \label{eq:dijetbound2}
\frac{\big | \hspace{0.125mm} C_{bG} \hspace{0.125mm} \big |}{\Lambda^2} < \left ( \frac{1}{1.6 \, {\rm TeV}} \right )^2 \,.
\eeq
This bound is by a factor of about 2.5 better than the estimate that has been given in~\cite{Bramante:2014hua} based on  the CMS dijet measurement~\cite{CMS-PAS-EXO-12-023}. Let us remark that adding an additional uncertainty of 10\% to the background prediction obtained in~\cite{Aad:2019hjw} would lead to a slightly weaker bound than~(\ref{eq:dijetbound2}) of approximately $1/(1.5 \, {\rm TeV})^2$. Since the unnormalised two $b$-tagged MC distribution and the data are found to agree within around 10\%  in the mass range $1 \, {\rm TeV} < M_{jj} <  5 \, {\rm TeV}$~\cite{Aaboud:2018tqo}, we believe that the latter bound is very robust.  Notice finally that searches for jet pairs with one or two jets identified as $c$-jets could be used to set a bound on the Wilson coefficient $| \hspace{0.125mm} C_{cG} \hspace{0.125mm} \big |$ similar to~(\ref{eq:dijetbound2}). To our knowledge dijet searches that employ $c$-tagging have however not been performed at the LHC to date.

\section{Constraints from $\bm{Z}$-boson production in association with $\bm{b}$-jets}
\label{sec:Zbjets}

Inclusive and differential measurements of $Z$-boson production in association with $b$-jets have been performed at the LHC by both the ATLAS and CMS collaborations~\cite{Chatrchyan:2013zja,Chatrchyan:2014dha,Aad:2014dvb,Khachatryan:2016iob,Sirunyan:2020lgh,Aad:2020gfi,ATLAS-CONF-2021-012}. These measurements provide not only an important role in improving our quantitative understanding of perturbative~QCD~(see~e.g.~\cite{Gauld:2020deh}), but can also be exploited to search for~BSM~physics~\cite{Afik:2019htr}. In this section we will consider the measurements performed in~\cite{Aad:2020gfi} to set bounds on the Wilson coefficient of the operator $Q_{bG}$. 

\begin{figure}[!t]
\begin{center}
\begin{subfigure}[c]{0.425\linewidth}
      \centering
       \scalebox{1.2}{
      \begin{tikzpicture}
      \begin{feynman}
      \vertex (a1) {\(g\)};
       \vertex[below=1.6cm of a1] (a2){\(g\)};
       \vertex[right=1.5cm of a1](a3);
         \vertex[below=1.6cm of a3](a4); 
        \vertex[right=1.5cm of a3](a5);
         \vertex[above=0.5cm of a5](a52){\(b\)};
        \vertex[right=1.5cm of a4](a6);
         \vertex[below=0.5cm of a6](a62){\(\bar b\)};
        \vertex[below=0.8cm of a3](a7);
        \vertex[right=1.2cm of a7](a8);
        \vertex[right=1cm of a8](a9);
        \vertex[above=0.4cm of a9](a10){\( \ell^+\)};
        \vertex[below=0.4cm of a9](a11){\( \ell^-\)};
       \diagram* {
        (a1) -- [gluon] (a3) -- [anti fermion] (a7)--[anti fermion] (a4) -- [gluon](a2), 
        (a3) -- [fermion](a52),
        (a4) -- [anti fermion]  (a62),
        (a7) -- [boson, edge label=\(Z\)] (a8),
        (a10) -- [ fermion] (a8) -- [ fermion] (a11);
                 };    
        \vertex[dot,fill=black] (d) at (a4){};
        \vertex[dot,fill=black] (d) at (a7){};
         \vertex[dot,fill=black] (d) at (a8){};
       \vertex[square dot,fill=nicered] (d) at (a3){};
      \end{feynman}
      \end{tikzpicture}
      }
\end{subfigure}%
\begin{subfigure}[c]{0.525\linewidth}
       \centering
        \scalebox{1.2}{
       \begin{tikzpicture}
       \begin{feynman}
      \vertex (a1) {\(q\)};
       \vertex[below=2.5cm of a1] (a2){\(\bar q\)};
       \vertex[below=1.25cm of a1] (a3);
       \vertex[right=1.3cm of a3] (a4);
       \vertex[right=1.3cm of a4] (a5);
       \vertex[right=4.3cm of a1](a6) {\(b\)};
       \vertex[right=4.3cm of a2] (a7){\(\bar b\)}; 
       \vertex[right=.85cm of a5](a8);
       \vertex[above=0.625cm of a8](a9);
       \vertex[right=0.85cm of a9](a10);
       \vertex[below=0.45cm of a10](a11);
        \vertex[right=1cm of a11](a12);
        \vertex[above=0.4cm of a12](a13){\( \ell^+\)};
        \vertex[below=0.4cm of a12](a14){\( \ell^-\)};
       \diagram* {
           (a1)--[fermion] (a4)-- [fermion] (a2),
           (a4)--[gluon, edge label=\(g\)] (a5),
           (a6)--[anti fermion] (a9) --[anti fermion] (a5)--[anti fermion] (a7),
           (a9) --[boson, edge label'=\( Z\)] (a11),
           (a13)--[fermion] (a11) --[fermion] (a14),
         };
          \vertex[dot,fill=black] (d) at (a4){};
       \vertex[square dot,fill=nicered] (d) at (a5){};
        \vertex[dot,fill=black] (d) at (a9){};
         \vertex[dot,fill=black] (d) at (a11){};
      \end{feynman}
       \end{tikzpicture}
       }
\end{subfigure}
\vspace{0mm}
\caption{\label{fig:Zbjets} Examples of Feynman diagrams leading to a $b \bar b \hspace{0.5mm} \ell^+ \ell^-$ signature at the LHC. The red squares denote the insertion of the bottom-quark chromomagnetic dipole moment. Graphs with a virtual photon $\gamma^\ast$ also contribute but are not shown explicitly. } 
\end{center}
\end{figure}
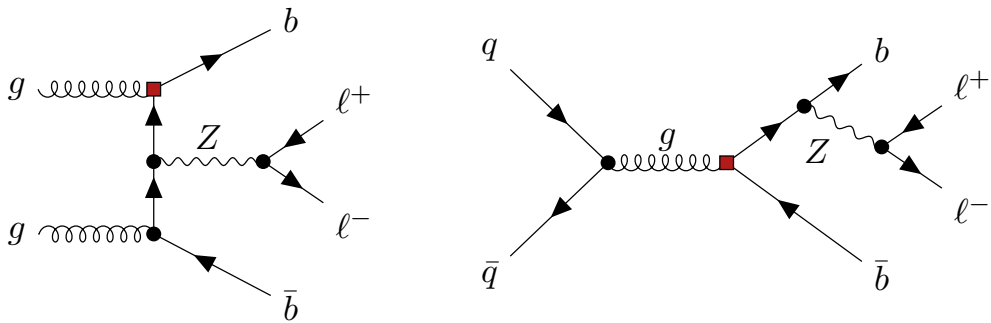

We calculate the contribution of the operator $Q_{bG}$ to $pp \to b \bar b  \hspace{0.5mm}  Z/\gamma^\ast \to  b \bar b \hspace{0.5mm}  \ell^+ \ell^-$ at LO using  {\tt MadGraph5\_aMC\@NLO}  together with  {\tt PYTHIA~8.2} and  {\tt NNPDF31\_nlo\_as\_0118} PDFs. The computation is performed in the four-flavour scheme. Relevant Feynman diagrams are shown in Figure~\ref{fig:Zbjets}.  
 To be able compare our results to an existing analysis we employ the event selections of the recent ATLAS measurement~\cite{Aad:2020gfi}  which is  based on $35.6 \, {\rm fb}^{-1}$ of $13 \, {\rm TeV}$~LHC data. Electrons must satisfy the tight likelihood requirement~\cite{Aaboud:2019ynx}, and are required to have a transverse momentum and pseudorapidity of  $p_T (e) > 27 \, {\rm GeV}$ and $|\eta (e)| < 2.47$, respectively. Electron candidates in the transition region between the barrel and endcap electromagnetic calorimeters,~i.e.~$1.37 <  |\eta(e)|  < 1.52$, are excluded. Muons 
 must satisfy medium identification criteria~\cite{Aad:2016jkr} and have to pass the requirements $p_T (\mu) > 27 \, {\rm GeV}$ and $|\eta (\mu)| < 2.5$. Jets are reconstructed using the anti-$k_t$ algorithm with radius parameter $R=0.4$, and a  $b$-tagging algorithm is applied with  a  $b$-tagging efficiency of 70\% and misidentification rates of 8.3\% of 0.26\% for $c$-jets and a light-flavour jets, respectively.  Events are required to have exactly two same-flavour ($\ell = e, \mu$) opposite-sign leptons  with a dilepton invariant mass in the range $76 \, {\rm GeV} < M_{\ell^+ \ell^-} < 106 \, {\rm GeV}$. To suppress the background from $t \bar t$ events with dileptonic decays, events with $p_T (\ell^+ \ell^-) < 150 \, {\rm GeV}$ must also have  missing transverse energy of $E_T^{\rm miss} < 60 \, {\rm GeV}$. Events passing the above selection are then assigned to two signal regions depending on the number of identified $b$-jets, and we concentrate in the following  on the signal region with at least two $b$-jets. The object and event selection used in our analysis are implemented into {\tt CheckMATE~2} which uses  {\tt DELPHES~3} as a fast detector simulation. 
 
\begin{figure}[!t]
\begin{center}
\includegraphics[width=0.475\textwidth]{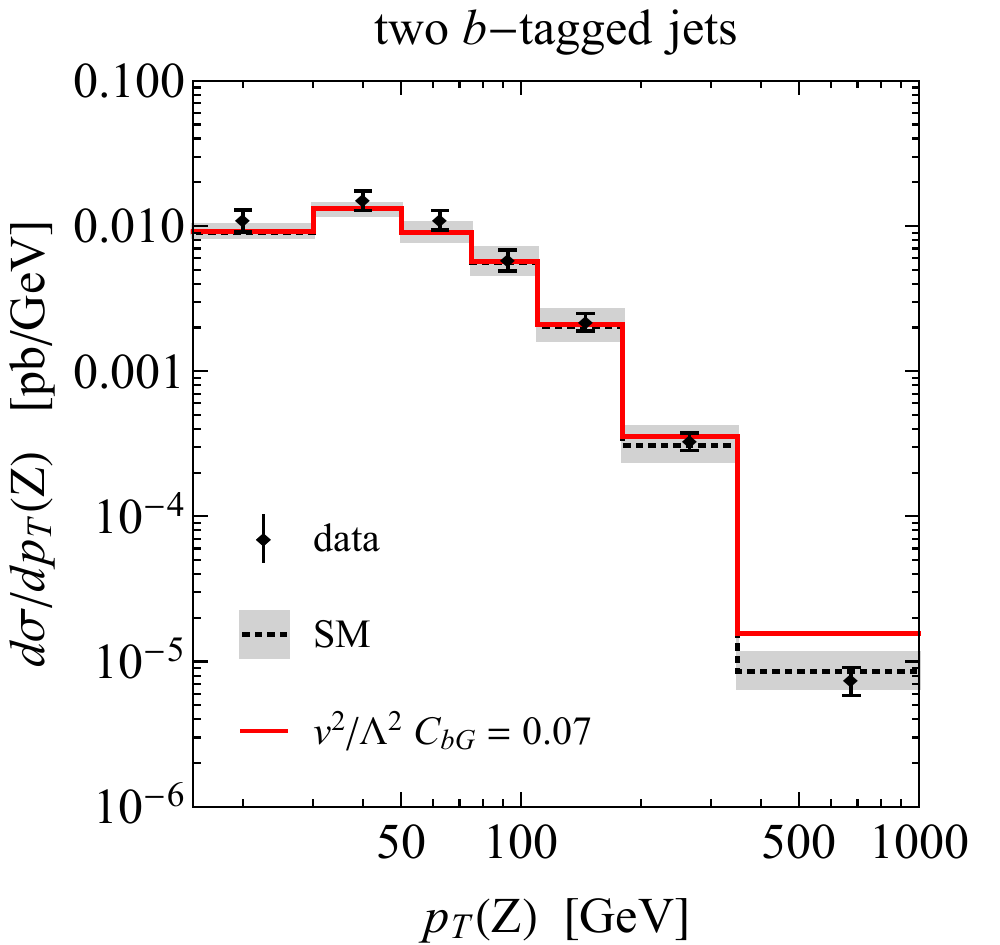} \quad 
\includegraphics[width=0.445\textwidth]{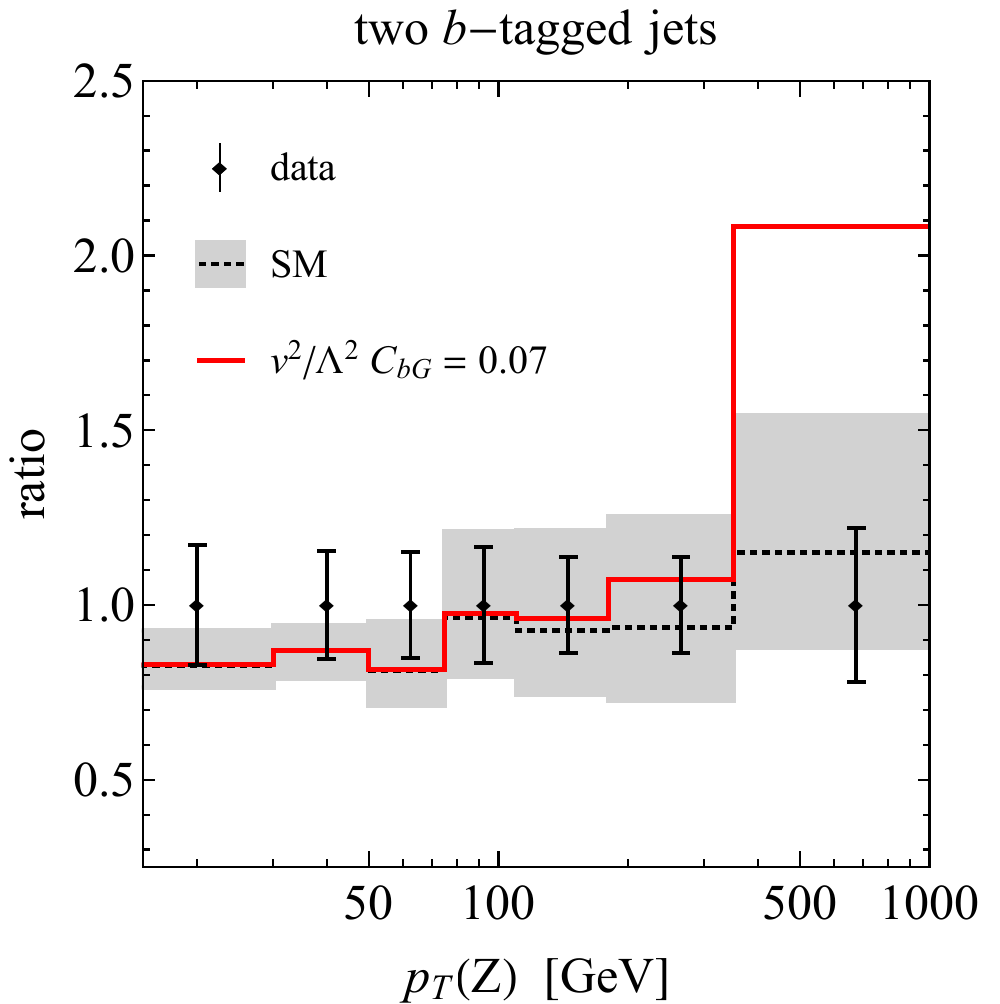}
\vspace{2mm} 
\caption{\label{fig:bbllplots} Left: $p_{T} (Z)$ distributions for the category with two $b$-jets. The~black dotted line is the central value of the SM prediction provided by ATLAS~\cite{Aad:2020gfi}, the  grey band indicates the associated uncertainties and the red curve shows the new-physics prediction assuming $v^2/\Lambda^2 \hspace{0.5mm} C_{bG} = 0.07$. Right:~Ratio between the  SM and the BSM prediction for $v^2/\Lambda^2 \hspace{0.5mm} C_{bG} = 0.07$ and the measurement. The colour coding resembles the one employed on the left-hand side.} 
\end{center}
\end{figure}

Several differential observables for the $pp \to b \bar b  \hspace{0.5mm}  Z/\gamma^\ast \to  b \bar b \hspace{0.5mm}  \ell^+ \ell^-$ process have been considered in the article~\cite{Aad:2020gfi}. Here we   focus on the transverse momentum of the $Z$-boson~$\big($$p_{T} (Z)$$\big)$. In~Figure~\ref{fig:bbllplots} we display results for the $p_{T} (Z)$ spectrum requiring two or more $b$-jets. The black dotted line and the grey band  in the left panel are the SM prediction with its uncertainty, while the red curve shows our new-physics prediction assuming~$v^2/\Lambda^2 \hspace{0.5mm} C_{bG} = 0.07$. The used SM numbers correspond to those referred to as  Sherpa~5FNS~(NLO) in~\cite{Aad:2020gfi}. They have been  obtained by ATLAS using~the~{\tt Sherpa~v2.2.1} generator~\cite{Bothmann:2019yzt}. From the two panels it is evident that the relative effect of a non-zero contribution due to $Q_{bG}$ grows with~ transverse momentum and as a result the BSM~prediction differs most visibly in the highest bin with $p_{T} (Z) > 350 \, {\rm GeV}$. 

Considering all seven bins of the ATLAS study~\cite{Aad:2020gfi} in a $\Delta \chi^2$ fit, we obtain 
\beq \label{eq:bbllbound}
\frac{\big | \hspace{0.125mm} C_{bG} \hspace{0.125mm} \big |}{\Lambda^2} < \left ( \frac{1}{980 \, {\rm GeV}} \right )^2 \,, 
\eeq
at  95\%~CL. We stress that this bound depends in a notable fashion on the systematic experimental and theoretical uncertainties that plague $pp \to  b \bar b  \hspace{0.5mm}  Z/\gamma^\ast \to  b \bar b \hspace{0.5mm}  \ell^+ \ell^-$ at high~$p_{T} (Z)$. In the  bin with $p_{T} (Z)> 350 \, {\rm GeV}$ the experimental (theoretical) uncertainty used in our analysis amounts to about~20\%~(35\%). Reducing the theoretical  uncertainty  by a factor of 2 would improve the limit~(\ref{eq:bbllbound}) to approximately  $1/(1.1 \, {\rm TeV} )^2$. In view of the recent progress~\cite{Gauld:2020deh} 
in the calculation of $Z$-boson production  in association with $b$-jets at next-to-next-to-leading order accuracy in QCD including finite heavy-quark mass effects, such a reduction of uncertainties does not seem unreasonable. 

\section{Constraints from flavour physics}
\label{sec:flavour}

In order to derive a bound on the real part of the Wilson coefficients of the charm-quark dipole operator, let us consider the following effective interaction
\beq \label{eq:Leffcharm}
{\cal L}_{\rm eff} \supset -\tilde \mu_c (m_h)  \hspace{0.5mm} \frac{ g_s(m_h)}{2} \hspace{0.5mm}  \bar c\hspace{0.25mm}  \sigma_{\mu \nu} T^a  \hspace{0.25mm} c  \hspace{0.5mm} G^{a, \mu \nu} \,,  
\eeq
where the initial condition $\tilde \mu_c (m_h)$ of the charm-quark chromomagnetic dipole moment in the terms of the relevant Wilson coefficient multiplying the operators in (\ref{eq:dipoleoperators}) is given  by 
\beq \label{eq:mutildec}
\tilde \mu_c (m_h) = - \frac{\sqrt{2} \hspace{0.25mm} v}{\Lambda^2} \hspace{0.5mm} {\rm Re} \hspace{0.25mm} \big ( C_{cG} \big ) \,.
\eeq
One-loop Feynman diagrams involving a $W$-boson exchange generate the  chromomagnetic dipole operator
\beq \label{eq:Q8}
Q_8 = \frac{g_s}{(4 \pi)^2} \hspace{0.25mm} m_b  \hspace{0.25mm}  \bar s_L \hspace{0.25mm}  \sigma_{\mu \nu} T^a \hspace{0.25mm}  b_R \hspace{0.5mm}  G^{a, \mu \nu} \,, 
\eeq
which appears in the $\Delta B = 1$ Lagrangian ${\cal L}  = -4 \hspace{0.25mm} G_F/\sqrt{2} \hspace{0.5mm} V_{ts}^\ast V_{tb} \hspace{0.25mm}  C_8 \hspace{0.25mm}Q_8$ with $G_F \simeq 1/\big (\sqrt{2} \hspace{0.25mm} v^2 \big)$ the Fermi constant  and  $V_{ij}$ the elements of the Cabibbo-Kobayashi-Maskawa~(CKM) matrix. The corresponding matching correction reads~\cite{Sala:2013osa,Gorbahn:2014sha}
\beq \label{eq:DC8}
\delta C_8 (m_h) = \frac{\bar m_c (m_h)}{2} \frac{V_{cs}^\ast V_{cb}}{V_{ts}^\ast V_{tb}} \, \tilde \mu_c (m_h) \,,
\eeq
where $\bar m_c (m_h)$ denotes  the charm-quark $\overline{\rm MS}$ mass evaluated at the Higgs-boson mass threshold. In terms of the shift $\delta C_8 (m_h)$, the ratio between the  branching ratio of $B \to X_s \gamma$ and its SM prediction can be written as~\cite{Misiak:2015xwa}
\beq \label{eq:RXsformula}
R_{X_s} = \frac{{\rm BR} \left ( B \to X_s \gamma \right ) }{{\rm BR} \left ( B \to X_s \gamma   \right )_{\rm SM}} = 1 - 0.59  \hspace{0.25mm} \delta C_8  (m_h)\,.
\eeq
Combining the SM calculation of  $B \to X_s \gamma$  with the present world average, one gets~\cite{Misiak:2017bgg}
\beq  \label{eq:RXsexp}
R_{X_s}= 0.97 \pm 0.08 \,,
\eeq
if uncertainties are added in quadrature. Using now that $V_{cs}^\ast V_{cb} \simeq - V_{ts}^\ast V_{tb}$ to high accuracy and employing $m_c (m_h) \simeq 0.87 \, {\rm GeV}$, one obtains from~(\ref{eq:mutildec}) to (\ref{eq:RXsexp}) the following 95\%~CL limit
\beq \label{eq:flavourbound}
\frac{\left | \hspace{0.25mm} {\rm Re} \hspace{0.25mm} \big ( C_{cG} \big )  \right |}{\Lambda^2} < \left ( \frac{1}{25 \, {\rm GeV}} \right )^2 \,.
\eeq
We add that a bound that is weaker than~(\ref{eq:flavourbound}) by a factor of around 4 is obtained by confronting the SM prediction  ${\rm BR} \left ( B \to X_s g\right )_{\rm SM} \simeq 5 \cdot 10^{-3}$~\cite{Greub:2000an,Greub:2000sy} with the corresponding  experimental limit that reads ${\rm BR} \left ( B \to X_s g\right ) \lesssim 10\%$ \cite{Kagan:1997sg}.   An even weaker bound follows from the isospin asymmetry in $B \to K^\ast \gamma$~\cite{Dimou:2012un,Lyon:2013gba,HFLAV:2019otj}. The~above estimate shows clearly that indirect probes of the chromodipole operators through flavour-physics observables cannot compete with the constraints arising from dijets. Notice that this is particularly true for the case of $Q_{bG}$ where a matching correction to $Q_8$ first arises at the two-loop level leading to a further suppression of~(\ref{eq:DC8}) by a factor of the order of $(4 \pi)^2$. We finally remark that CP~violation in the $\Delta C =1 $ sector --- for instance~in the form of the difference $\Delta A_{\rm CP}$ between the two direct CP~asymmetries in $D \to K^+ K^-$ and $D \to \pi^+ \pi^-$ --- also does not provide relevant bounds on ${\rm Im} \hspace{0.25mm} \big ( C_{bG} \big )$ due to the strong CKM and loop suppression.  

\section{Constraints from nEDM}
\label{sec:nEDM}

Searches for EDMs are known to place stringent constraints on any new-physics scenario with additional sources of CP violation (see~\cite{Pospelov:2005pr,Li:2010ax,Kamenik:2011dk,McKeen:2012av,Chang:2013cia,Engel:2013lsa,Jung:2013hka,Gripaios:2013lea,Brod:2013cka,Sala:2013osa,Inoue:2014nva,Gorbahn:2014sha,Altmannshofer:2015qra,Dwivedi:2015nta,Chien:2015xha,Cirigliano:2016njn,Cirigliano:2016nyn,Dekens:2018bci,Cesarotti:2018huy,Panico:2018hal,Brod:2018pli,Brod:2018lbf,Cirigliano:2019vfc,Haisch:2019xyi} for reviews and recent discussions). To~derive a bound on the imaginary part of the Wilson coefficient of the bottom-quark dipole operator appearing  in~(\ref{eq:dipoleoperators}), we consider the following effective interactions
\beq \label{eq:Leff}
{\cal L}_{\rm eff} \supset -\tilde d_b (m_h)  \hspace{0.5mm} \frac{i \hspace{0.25mm} g_s(m_h)}{2} \hspace{0.5mm}  \bar b\hspace{0.25mm}  \sigma_{\mu \nu} T^a \gamma_5 \hspace{0.25mm} b  \hspace{0.5mm} G^{a, \mu \nu} - w(m_h) \hspace{0.5mm}  \frac{1}{3} \hspace{0.25mm}  f^{abc} \hspace{0.25mm} G_{\mu \sigma}^a  \hspace{0.25mm}  G_{\nu}^{b, \sigma}  \hspace{0.25mm}  \widetilde{G}^{c, \mu \nu}   \,,  
\eeq
where $\widetilde G^{a,\mu \nu} = 1/ 2 \hspace{0.5mm} \epsilon^{ \hspace{0.25mm} \mu \nu \alpha \beta}  \hspace{0.25mm}  G_{\alpha \beta}^a$ is the dual field-strength tensor of QCD with $ \epsilon^{ \hspace{0.25mm} \mu \nu \alpha \beta} $ the fully antisymmetric Levi-Civita tensor ($\epsilon^{0123} = 1$). 

In the presence of the two dimension-six SMEFT operators~(\ref{eq:dipoleoperators}) the initial condition  $\tilde d_b (m_h)$ of the bottom-quark chromoelectric dipole moment is given by 
\beq \label{eq:dtildeb}
\tilde d_b (m_h) = - \frac{\sqrt{2} \hspace{0.25mm} v}{\Lambda^2} \hspace{0.5mm} {\rm Im} \hspace{0.25mm} \big ( C_{bG} \big ) \,, 
\eeq
while the initial condition  $w (m_h)$ of the Weinberg operator~\cite{Weinberg:1989dx} vanishes. The  bottom-quark chromoelectric dipole moment has the following  QCD running 
\beq \label{eq:dtildeRGE}
\tilde d_b (m_b) = \left ( \frac{\alpha_s (m_h)}{\alpha_s (m_b)} \right )^{\frac{\gamma_{\tilde d}}{2 \gamma_{\alpha_s}}} \hspace{0.5mm} \tilde d_b (m_h) \,,
\eeq
where $\gamma_{\alpha_s}$ denotes the LO anomalous dimension (or beta function) of the strong coupling constant and $ \gamma_{\tilde d}$ is the LO anomalous dimension of the bottom-quark chromoelectric dipole moment. Explicitly, one has 
\beq \label{eq:gammaasgammabG}
\gamma_{\alpha_s} =  11 - \frac{2}{3} \hspace{0.25mm} N_f \,, \qquad \gamma_{\tilde d} = \frac{4}{3} \,,
\eeq
where the expression for  $\gamma_{\alpha_s}$ can be found in any textbook on QCD and the result for $\gamma_{\tilde d}$ has  first been presented in~\cite{Shifman:1976de}. Notice that in~(\ref{eq:dtildeRGE}) the number of active flavours is $N_f = 5$ and that one could in principle resum large logarithms in the latter equation up to the next-to-next-to-leading logarithmic level because the self-mixing of the dipole operators  is known up to the three-loop order~\cite{Gorbahn:2005sa}. Given the sizeable uncertainties of the hadronic matrix elements of the  Weinberg operator and the chromoelectric dipole operators of light quarks $\big($cf.~the discussion before~(\ref{eq:NEDMformulab1})$\big)$ using only the leading-logarithmic~(LL) result~(\ref{eq:dtildeRGE}) is however fully~justified.

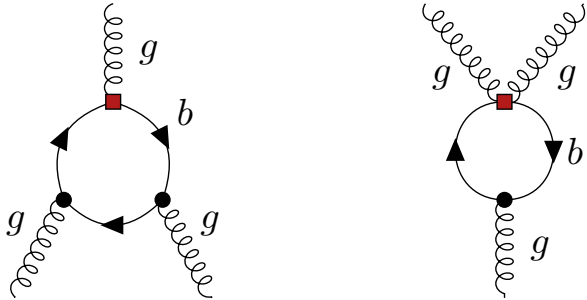
\begin{figure}[!t]
\begin{center}
\begin{subfigure}[b]{0.4\linewidth}
     \begin{flushright}
     \scalebox{1.3}{
      \begin{tikzpicture}
      \begin{feynman}
      \vertex (a1);
      \vertex[below=1cm of a1](a2);      
      \vertex[below=1cm of a2](a3);
       \vertex[right =0.5cm of a3](a4);
       \vertex[left =0.5cm of a3](a5);     
       \vertex[below =1cm of a3](a6);
        \vertex[right =1cm of a6](a7);
       \vertex[left =1cm of a6](a8);         
       \diagram* {
       (a1) -- [gluon, edge label=\( \ g\)] (a2),
       (a2) -- [quarter left, looseness=1.25, fermion, edge label= \(b \)] (a4) -- [quarter left, looseness=1.25, fermion](a5) -- [quarter left, looseness=1.25, fermion](a2),
       (a4) -- [gluon, edge label=\(g\)](a7),
       (a5) -- [gluon, edge label'=\(g\)](a8),
              };
        \vertex[dot,fill=black] (d) at (a4){};
       \vertex[dot,fill=black] (d) at (a5){};
       \vertex[square dot,fill=nicered] (d) at (a2){};
      \end{feynman}
      \end{tikzpicture}
      }
      \end{flushright}
\end{subfigure}%
\hspace{2cm}                                                                                                                                                                     
\begin{subfigure}[b]{0.4\linewidth}
      \scalebox{1.3}{
       \begin{tikzpicture}
      \begin{feynman}
      \vertex (a1) ;
      \vertex[left=0.8cm of a1](a11);
      \vertex[right=0.8cm of a1](a12);
      \vertex[below=1cm of a1](a2);
      \vertex[below=1cm of a2](a3);
      \vertex[below =1cm of a3](a4);
       \diagram* {
       (a11) -- [gluon, edge label'=\(g\)] (a2) -- [gluon, edge label'=\(g\)] (a12),
       (a2) -- [half left, looseness=1.7, fermion, edge label= \(b \)] (a3) -- [half left, looseness=1.7, fermion](a2),
       (a3) -- [gluon, edge label=\( \ g\)](a4),
              };
      \vertex[dot,fill=black] (d) at (a3){};
       \vertex[square dot,fill=nicered] (d) at (a2){};
      \end{feynman}
       \end{tikzpicture}
       }
\end{subfigure}
\vspace{2mm}
\caption{\label{fig:bottomthreshold} One-loop diagrams leading to a correction to the Weinberg operator at the bottom-quark threshold. The red squares denote the insertion of the bottom-quark chromoelectric dipole moment. } 
\end{center}
\end{figure}

At the bottom-quark mass threshold one integrates out the bottom quark. The corresponding Feynman diagrams are shown in Figure~\ref{fig:bottomthreshold}. This gives a finite matching correction to the Weinberg operator~\cite{Boyd:1990bx,Braaten:1990gq,Chang:1990jv}
\beq \label{eq:weinbergthreshold}
\delta w (m_b) = \frac{g_s^3 (m_b)}{32 \pi^2} \frac{\tilde d_b (m_b) }{\bar m_b (m_b)}  \,.
\eeq
Between the bottom-mass scale and the hadronic scale $\mu_H \simeq 1 \, {\rm GeV}$ the Weinberg operator mixes into itself and into the chromoelectric and  electric dipole operators involving up and down quarks. As it turns out, the contributions from  electric dipole operators are numerically very small and we thus neglect it in what follows.  The   LO anomalous dimensions  describing the self-mixing of the Weinberg operator and the mixing of the Weinberg operator in the chromoelectric dipole operators, respectively,  are~\cite{Braaten:1990gq,Braaten:1990zt,deVries:2019nsu} 
\beq \label{eq:ADMw}
\gamma_{w} = 3 +2  N_f\,, \qquad \gamma_{w\tilde d } = - 6 \,.
\eeq
Resumming LL corrections in the four- and the three-flavour theory, we obtain
\beq \label{eq:wdqb}
w(\mu_H) \simeq 0.71 \hspace{0.25mm} \delta w (m_b) \,, \qquad 
\tilde d_q (\mu_H) \simeq  0.14 \hspace{0.5mm} \bar m_q (\mu_H) \hspace{0.5mm}  \delta w (m_b) \,,
\eeq
where $q = u, d$. The numerical factors in (\ref{eq:wdqb}) correspond to the following values $\alpha_s(m_b) \simeq 0.21$,  $\alpha_s(m_c) \simeq 0.32$ and  $\alpha_s(\mu_H) \simeq 0.36$ of the QCD coupling constant.

Estimates of the hadronic matrix elements of the Weinberg operator and the chromomagnetic dipole operators of the up and down quark are  presently plagued by uncertainties of 50\% --- see for instance~\cite{Pospelov:2000bw,Demir:2002gg,Lebedev:2004va,Hisano:2012sc,Pospelov:2005pr,Haisch:2019bml,Yamanaka:2020kjo,Hatta:2020riw}. Adopting the QCD sum-rule estimates~\cite{Pospelov:2000bw,Haisch:2019bml} for the relevant hadronic matrix elements and employing  $\bar m_u(\mu_H) \simeq 2.5 \cdot 10^{-3} \, {\rm GeV}$ and $\bar m_d(\mu_H) \simeq 5.4 \cdot 10^{-3} \, {\rm GeV}$~\cite{Zyla:2020zbs}, the nEDM can  be written as
\beq \label{eq:NEDMformulab1}
\frac{d_n}{e} \simeq \big (  -1.77 \left ( 1 \pm 0.5 \right ) + 0.10 \left ( 1 \pm 0.5 \right ) \big  ) \cdot 10^{-2} \, {\rm GeV} \hspace{0.5mm} \delta w (m_b) \,,
\eeq
where the first and the second  term corresponds to the contribution from the Weinberg operator and the chromomagnetic dipole operators, respectively.  Combining  the above uncertainties  in such a way that our prediction provides a lower absolute limit on the actual size of the $Q_{bG}$ correction to the nEDM, we find 
\beq \label{eq:NEDMformulab2}
\left | \frac{d_n}{e}  \right |  \simeq 1.5 \cdot 10^{-22} \, {\rm cm} \, \left ( \frac{1 \, {\rm TeV}}{\Lambda} \right )^2 \, \left |\hspace{0.25mm}  {\rm Im} \hspace{0.25mm} \big ( C_{bG} \big ) \right |  \,. 
\eeq
Here we have employed $\alpha_s (m_h) \simeq 0.11$ and $\bar m_b (m_b) \simeq 4.2 \, {\rm GeV}$~\cite{Zyla:2020zbs} to obtain the numerical result.  The current best experimental nEDM  result~\cite{Abel:2020gbr} imposes the following 90\%~CL bound
\beq \label{eq:NEDMexp}
\left | \frac{d_n}{e} \right | < 1.8 \cdot 10^{-26} \, {\rm cm} \,.
\eeq
From~(\ref{eq:NEDMformulab2}) and~(\ref{eq:NEDMexp}) it follows that 
\beq \label{eq:dipoleboundb}
\frac{\left | \hspace{0.25mm} {\rm Im} \hspace{0.25mm} \big ( C_{bG} \big ) \hspace{0.25mm}  \right |}{\Lambda^2} < \left ( \frac{1}{90 \, {\rm TeV}} \right )^2 \,.
\eeq 

In the case of the  charm-quark dipole operator entering~(\ref{eq:dipoleoperators}), the derivation of the bound on the imaginary part of $C_{cG}$ follows the steps described above with minor modifications.  The first modification is that from the initial scale $m_h$ the charm-quark chromomagnetic dipole moment $\tilde d_c$ runs to the charm-quark threshold $m_c$ in the five- and four-flavour theory.  Second, the renormalisation group flow below $m_c$ where the Weinberg operator is generated --- see (\ref{eq:weinbergthreshold}) ---  proceeds only in the three-flavour theory, leading to 
\beq \label{eq:wdqc}
w(\mu_H) \simeq 0.94 \hspace{0.5mm} \delta w (m_c) \,, \qquad 
\tilde d_q (\mu_H) \simeq  0.04 \hspace{0.5mm} \bar m_q (\mu_H) \hspace{0.5mm}  \delta w (m_c) \,.
\eeq
Notice that given the smaller scale separation between $m_c$ and $\mu_H$ compared to $m_b$ and $\mu_H$ the renormalisation group~effects in~(\ref{eq:wdqc}) are notably smaller than those in~(\ref{eq:wdqb}).  Instead of~(\ref{eq:NEDMformulab1}) one then finds the following expression 
\beq \label{eq:NEDMformulac1}
\frac{d_n}{e} \simeq \big (  -2.34 \left ( 1 \pm 0.5 \right ) + 0.03 \left ( 1 \pm 0.5 \right ) \big  ) \cdot 10^{-2} \, {\rm GeV} \hspace{0.5mm} \delta w (m_c) \,,
\eeq
which is fully dominated by the contribution from the Weinberg operator. This leads to 
\beq \label{eq:NEDMformulac2}
\left | \frac{d_n}{e}  \right |  \simeq 1.4 \cdot 10^{-21} \, {\rm cm} \, \left ( \frac{1 \, {\rm TeV}}{\Lambda} \right )^2 \, \left | \hspace{0.25mm} {\rm Im} \hspace{0.25mm} \big ( C_{cG} \big )\right |  \,, 
\eeq
where we have used $\bar m_c (m_c) \simeq 1.3 \, {\rm GeV}$~\cite{Zyla:2020zbs}. Combining~(\ref{eq:NEDMexp}) and~(\ref{eq:NEDMformulac2}) we finally obtain 
\beq \label{eq:dipoleboundc}
\frac{\left | \hspace{0.25mm} {\rm Im} \hspace{0.25mm} \big ( C_{cG} \big ) \hspace{0.25mm}  \right |}{\Lambda^2} < \left ( \frac{1}{275 \, {\rm TeV}} \right )^2 \,.
\eeq 

\section{Discussion}
\label{sec:discussion}

In Table~\ref{tab:summary} we summarise the limits on the Wilson coefficients of the operators $Q_{bG}$ and $Q_{cG}$ that we have derived in the course of this work. For what concerns the magnitudes of $C_{bG}$ and $C_{cG}$, one sees that searches for dijets final states provide bounds on the Wilson coefficients of order one assuming a suppression scale $\Lambda$  of $1 \, {\rm TeV}$.  In fact, the nominal strongest limit derives at present from the ATLAS search for $b$-jet pairs performed  in~\cite{Aad:2019hjw}. This search probes dijet invariant masses in the range $1133 \, {\rm GeV} < M_{jj} <  4595 \, {\rm GeV}$, which pushes the  Wilson coefficient to $\big | \hspace{0.125mm} C_{bG} \hspace{0.125mm} \big | \gtrsim 4 \pi$, if one requires~$\Lambda \gtrsim 5 \, {\rm TeV}$ to cover the whole range of tested jet-jet invariant masses. Similar~statements apply to the constraints following from the CMS measurement of the dijet angular distributions~\cite{Sirunyan:2018wcm}. On general grounds, the limits~(\ref{eq:dijetbound1}) and (\ref{eq:dijetbound2}) therefore only constrain strongly-coupled  ultraviolet~(UV) completions --- for related discussions of the applicability of the SMEFT to dijet searches see~\cite{Krauss:2016ely,Alioli:2017jdo,Alte:2017pme,Hirschi:2018etq,Keilmann:2019cbp,Goldouzian:2020wdq}. The situation is better in the case of the bound on~$C_{bG}$ that we have derived from the recent ATLAS measurement~\cite{Aad:2020gfi} of $pp \to b \bar b  \hspace{0.5mm}   Z/\gamma^\ast \to b \bar b  \hspace{0.5mm}   \ell^+ \ell^-$,  because the energy scales tested by this measurement are all below $1 \, {\rm TeV}$. Notice  that $pp \to b \bar b  \hspace{0.5mm}   Z/\gamma^\ast \to b \bar b  \hspace{0.5mm}   \ell^+ \ell^-$ measurements are systematically limited at the LHC but the prospects of reducing the systematic uncertainties due to an improved theoretical understanding are quite good~(cf.~\cite{Gauld:2020deh}). These two features taken together make $Z$-boson  production in association with $b$-jets  in our opinion a key process to search and to constrain the chromomagnetic bottom-quark dipole operator at upcoming~LHC~runs.  Another interesting avenue to  constrain chromomagnetic dipole operators involving the bottom quark  is HL-LHC Higgs physics~\cite{Hayreter:2013kba,Bramante:2014hua}, and we plan to return to this topic in future research. Furthermore, as~can be learnt  from Table~\ref{tab:summary} as well as the discussion at the end of Section~\ref{sec:flavour}, flavour physics does not provide a meaningful bound on either the operator $Q_{bG}$ or the operator $Q_{cG}$.   We finally add that HL-LHC constraints on   $\big | \hspace{0.125mm} C_{bG} \hspace{0.125mm} \big |$ from Higgs physics have been derived in~\cite{Hayreter:2013kba,Bramante:2014hua}. For instance, the article~\cite{Bramante:2014hua} obtained $\big | \hspace{0.125mm} C_{bG} \hspace{0.125mm} \big |/\Lambda^2 < 1/(2.1 \, {\rm TeV})^2$ by considering $pp \to b \bar b h \to  b \bar b \gamma \gamma$ production. 

\begin{table} 
\def\arraystretch{1.25}
\begin{center}
\begin{tabular}{cccc}
\hline 
Observable & Wilson coefficient & 95\%~CL bound& Scale \\[1mm]
\hline 
Dijet angular distributions & $\big | \hspace{0.125mm} C_{bG} \hspace{0.125mm} \big |$, $\big | \hspace{0.125mm} C_{cG} \hspace{0.125mm} \big |$  & $6.8$ & $380 \, {\rm GeV}$   \\[1mm]
Two $b$-tagged jets & $\big | \hspace{0.125mm} C_{bG} \hspace{0.125mm} \big |$ & $0.36$ & $1.6 \, {\rm TeV}$  \\[1mm]
$Z$-boson production with two $b$-jets  & $\big | \hspace{0.125mm} C_{bG} \hspace{0.125mm} \big |$ & $1.04$ & $980 \, {\rm GeV}$ \\[1mm]
Inclusive radiative $B$ decay & $\left |{\rm Re} \hspace{0.25mm} \big ( C_{cG} \big ) \right |$ & $1600$ & $25 \, {\rm GeV}$ \\[1mm]
Searches for nEDM  & $\left | {\rm Im} \hspace{0.25mm} \big ( C_{bG} \big  ) \right |$ & $1.2 \cdot 10^{-4}$ &  $90 \, {\rm TeV}$ \\[1mm]
Searches for nEDM  & $\left |{\rm Im} \hspace{0.25mm} \big (C_{cG} \big  ) \right |$ &$ 1.3 \cdot 10^{-5}$  & $275 \, {\rm TeV}$  \\[1mm]
\hline 
 \end{tabular}
\vspace{2mm}
\caption{Summary of the constraints on the Wilson coefficients of the operators $Q_{bG}$ and $Q_{cG}$ derived in this article. The given 95\%~CL bounds correspond to $\Lambda = 1 \, {\rm TeV}$, while the numbers quoted for the suppression scale assume  a purely real or imaginary Wilson coefficient with a magnitude of $1$. See text for further explanations.}
\label{tab:summary}
\end{center}
\end{table}

From Table~\ref{tab:summary} it is also clear that in contrast to $\big | \hspace{0.125mm} C_{bG} \hspace{0.125mm} \big |$ and $\big | \hspace{0.125mm} C_{cG} \hspace{0.125mm} \big |$, the imaginary parts of both Wilson coefficients are severely constrained by searches for a nEDM.  In the case of a purely imaginary Wilson coefficient $C_{bG}$ ($C_{cG}$) of $1$, scales as high as $90 \, {\rm TeV}$ ($275 \, {\rm TeV}$) are excluded at 90\%~CL by these indirect searches. Notice that  in weakly-coupled UV complete  theories the Wilson coefficients of chromodipole operators are typically both Yukawa- and loop-suppressed.  Assuming that $C_{bG} = y_b/(4 \pi)^2$  and $C_{cG} = y_c/(4 \pi)^2$ the bounds (\ref{eq:dipoleboundb}) and (\ref{eq:dipoleboundc}) imply $\Lambda > 960 \, {\rm GeV}$ and $\Lambda > 1.5 \, {\rm TeV}$, respectively. This numerical example shows that nEDM searches, unlike the studied collider constraints, are able to test weakly-coupled TeV-scale new-physics models. Making effects of bottom-quark and charm-quark chromodipole  operators observable at the LHC therefore  generically requires a mechanism that suppresses new sources of CP-violation beyond the SM. 

\acknowledgments 
UH thanks Darren~Scott, Marius~Wiesemann and Giulia~Zanderighi for triggering this work and Rhorry~Gauld for helpful correspondence concerning $pp \to b \bar b  \hspace{0.5mm}   Z/\gamma^\ast \to b \bar b  \hspace{0.5mm}   \ell^+ \ell^-$ production.  We are also grateful to Amando~Hala for reminding us of~\cite{Abel:2020gbr} and to Roman~Zwicky for his interest in our work.  All Feynman diagrams shown in this article have been drawn  with {\tt TikZ-Feynman}~\cite{Ellis:2016jkw}.


\providecommand{\href}[2]{#2}\begingroup\raggedright\endgroup

\end{document}